\documentclass[usenatbib]{mn2e}
\usepackage{graphicx}
\usepackage{amssymb}
\usepackage{lscape}
\usepackage{ulem}
\usepackage{txfonts}
\usepackage{url}
\usepackage{subfig}

\def\kms{km~s$^{-1}$}

\def\hi{H\,{\sc i}}

\def\kpc{$h_{70}^{-1}$ kpc}

\def\delv{$\Delta v$}

\def\delsfr{$\Delta$log(SFR)}

\newcommand{\rp}{$r_p$}
\def\fgas{$M_{gas}/M_*$}

\title[\hi~gas in galaxy pairs]{Galaxy Pairs in the Sloan Digital Sky Survey - X: Does gas content alter star formation rate enhancement in galaxy interactions?}

\author[Scudder et al.] {Jillian M. Scudder$^{1,2}$\thanks{J.Scudder@sussex.ac.uk}, Sara L. Ellison$^1$, Emmanuel Momjian$^3$, Jessica L. Rosenberg$^4$, 
\newauthor Paul Torrey$^{5,6,7}$, David R. Patton$^8$, Derek Fertig$^4$, J. Trevor Mendel$^9$\\
$^1$ Department of Physics and Astronomy, University of Victoria, Victoria, British Columbia, V8P 1A1, Canada.\\
$^2$ Astronomy Centre, Department of Physics \& Astronomy, University of Sussex, Brighton, BN1 9QH, England.\\
$^3$ NRAO, Domenici Science Operations Center, Socorro, NM, 87801, USA.\\
$^4$ School of Physics, Astronomy, and Computational Science, George Mason University, Fairfax, VA, 22030, USA.\\
$^5$ Harvard-Smithsonian Center for Astrophysics, 60 Garden Street, Cambridge, MA 02138, USA.\\
$^6$ Department of Physics, Kavli Institute for Astrophysics and Space Research, Massachusetts Institute of Technology, Cambridge, MA 02139, USA.\\
$^7$ TAPIR, Mailcode 350-17, California Institute of Technology, Pasadena, CA 91125, USA.\\
$^8$ Department of Physics \& Astronomy, Trent University, Peterborough, Ontario, K9J 7B8, Canada.\\
$^9$ Max-Planck-Institut f\"ur extraterrestrische Physik, 85748 Garching, Germany.}

\begin{document}

\maketitle

\begin{abstract}
New spectral line observations, obtained with the Jansky Very Large Array (VLA), of a sample of 34 galaxies in 17 close pairs are presented in this paper.  The sample of galaxy pairs is selected to contain galaxies in close, major interactions (i.e., projected separations $<$30 \kpc, and mass ratios less extreme than 4:1), while still having a sufficiently large angular separation that the VLA can spatially resolve both galaxies in the pair.  Of the 34 galaxies, 17 are detected at $> 3\sigma$.  We compare the \hi~gas fraction of the galaxies with the triggered star formation present in that galaxy. When compared to the star formation rates (SFRs) of non-pair galaxies matched in mass, redshift, and local environment, we find that the star formation enhancement is weakly positively correlated ($\sim 2.5\sigma$) with \hi~gas fraction.  In order to help understand the physical mechanisms driving this weak correlation, we also present results from a small suite of binary galaxy merger simulations with varying gas fractions. 
The simulated galaxies indicate that larger initial gas fractions are associated with lower levels of interaction-triggered star formation (relative to an identical galaxy in isolation), but also show that high gas fraction galaxies have higher absolute SFRs prior to an interaction.  
We show that when interaction-driven SFR enhancements are calculated relative to a galaxy with an average gas fraction for its stellar mass, the relationship between SFR and initial gas fraction dominates over the SFR enhancements driven by the interaction. 
Simulated galaxy interactions that are matched in stellar mass but not in gas fraction, like our VLA sample, yield the same general positive correlation between SFR enhancement and gas fraction that we observe.

\end{abstract}


\section{Introduction}
\label{sec:intro}
The interactions between galaxies are an ideal laboratory to study the response of galaxies to strong external perturbations.  
Galaxy-galaxy interactions very clearly drive strong changes in the perturbed galaxies.  
Interacting galaxies are universally observed to have higher than average star formation rates \citep[SFRs; e.g.,][]{Lambas2003, Alonso2004, Woods2007, Ellison2008, Ellison2010, Xu2010, Darg2010, Stierwalt2014}, diluted nuclear gas-phase metallicities \citep[e.g.,][]{Kewley2006, Michel-Dansac2008, Ellison2008, Scudder2012b}, bluer than average nuclear colours \citep[e.g.,][]{Patton2011}, and an overabundance of Active Galactic Nuclei \citep[AGN; e.g.,][]{Koss2011, Silverman2011, Ellison2011b, Sabater2013, Ellison2013, Satyapal2014b, Khabiboulline2014}.

The combination of observational results with the analysis of simulations of galaxy interactions allows us to interpret these data as the many indirect signs of gas flowing to the centre of the galaxy in response to the interaction.  
These gas flows are understood theoretically as the result of gravitationally driven tidal torques induced by the companion galaxy.  These torques act upon misaligned stellar and gaseous bars within the perturbed galaxies, produced as a result of the non-axisymmetric nature of the gravitational well, resulting from the presence of a companion galaxy.  The misalignment between the gaseous and stellar bars allows the stellar component to effectively remove angular momentum from the gaseous component, the loss of which causes the gas to fall towards the centre of the galaxy, producing the flow needed to explain the observations \citep[e.g.,][]{Mihos1994, Mihos1996, Barnes1996, Cox2006, diMatteo2007, Montuori2010, Torrey2012}. 
The gas inflow builds up a considerable reservoir of gas near the centre of the galaxy, which can then be efficiently converted into stars, fuelling a nuclear starburst \citep[{e.g.,}][]{diMatteo2007, Montuori2010, Torrey2012, Moreno2014}.
While galaxies in close pairs on average show higher than expected SFRs, the level of triggered star formation can vary widely from galaxy to galaxy, and from interaction to interaction
 \citep{Scudder2012b}.
Many previous studies have attempted to isolate parameters which govern the strength of the star formation triggered within a galaxy. A few parameters are consistently found to modulate a galaxy's response.

The mass ratio between the two galaxies permits the strongest SFR enhancements when the two galaxies are approximately of equal mass. Unequal mass mergers are observed to be much less effective at triggering strong starbursts \citep{Woods2007, Ellison2008, Xu2012, Lambas2012, Scudder2012b}, as the gravitational influence of a low mass companion should be weaker \citep{Cox2008,Ji2014}.
The environment of the pair can also suppress triggered SFR if the galaxy is found within a particularly dense region \citep[e.g.,][]{Alonso2004, Alonso2006, Baldry2006, Ellison2010}.
Observational studies have shown that the strongest enhancements in SFR are visible at the smallest projected separations \citep[\rp; e.g.,][]{Lambas2003, Alonso2004, Woods2007, Ellison2008, Scudder2012b, Patton2013}.  
Unlike the simulations, observational results are limited to a snapshot view of any individual interaction; 
\rp~serves as the best observational proxy for timescale and merger stage, properties which are predicted by theoretical studies to also modulate the SFR response of a galaxy interaction \citep[e.g.,][]{Torrey2012}.

Theoretical works have also offered several additional parameters which could alter the strength of a triggered starburst.
For instance, the increasing prominence of a galactic bulge should suppress triggered SFR \citep[e.g.,][]{diMatteo2007}.
Most simulations also agree that the orbital parameters imposed at the beginning of a simulation can have a significant effect upon the strength of the triggered star formation; the inclusion of a range of orbital parameters into a simulation suite introduces significant scatter.  \citep[e.g.,][]{Mihos1992, diMatteo2007, Cox2008, Perez2011a, Torrey2012}.  Generally, any configuration or orbit of the two galaxies which allows for a particularly strong tidal force to act upon the two galaxies will result in a stronger triggered starburst.  
However, the details of the orbital parameters of an interacting system are not readily constrained with observations, and therefore are generally left as a known source of scatter in the observations.  

The gas fraction of the interacting galaxies has also been suggested by the simulations to play a role in the triggering of SFR in galaxy pairs. However, the impact of gas fraction has been less well-developed theoretically than the impact of mass ratio or orbital inclination. 
    As such, theoretical works have not yet built a comprehensive framework to understand the impact of varying gas fraction on the SFRs of galaxies in pairs.  Different theoretical studies provide windows into the physical processes that may be at play.  Previous works \citep{Bournaud2011, Perez2011a, Perret2014} have investigated the SFR response in high redshift interactions, constructed to have high gas fractions. Galaxies with high gas fractions are found to sustain strong, often turbulent, star formation well before the two simulated galaxies interact.  
The interpretation of these results varies; strong enhancement in the SFR is described by \citet{Bournaud2011}, \citet{Perez2011a} found enhancement, but at a low level, and \citet{Perret2014} found no evidence for additional star formation due to the interaction.
\citet{Perret2014} suggests that at sufficiently high SFRs, the galaxy may be able to `saturate' its star formation, unable to drive SFR to higher values even though the interaction causes a significant perturbation.

However, these high redshift analogue simulations do not explicitly test the effect of varying the gas fraction within a galaxy on the SFR.  This is undertaken by \citet{Hopkins2009}, which suggests that the gas fraction of a galaxy is anti-correlated with its starburst efficiency.  This anti-correlation arises due to the physics behind the origin of the gas flows in interacting galaxies.
    If the merger's starburst strength is driven by the ability of the torque between the stellar and gaseous bar to efficiently strip the gaseous bar of angular momentum, the relative strengths of those bars may overwhelm the importance of the volume of gas available.  A high gas fraction galaxy would be very inefficient at driving gas flow, as a relatively weak stellar bar would exert a much smaller force upon a more massive gaseous bar.  The weaker force would not be able to drive large quantities of gas to central regions of the galaxy, and the nuclear starburst which followed would be marginal.  
A low gas fraction galaxy would show the inverse effect; a very efficient torque of the small amount of gas within the galaxy, an effect also suggested by \citet{Cox2008}.

In spite of the predicted relationship between increasing gas fraction and decreasing triggered star formation within the galaxy interaction, \citet{diMatteo2007} found that the orbital parameters of an encounter drove far more scatter in the amount of star formation triggered in an interaction than variations in the gas fraction did. The large scatter found by \citet{diMatteo2007} implies that any correlation between SFR enhancement and gas fraction may be buried due to the impact of including varying orientations.

\subsection{The current work}
To date, there is no comprehensive study of \hi~gas in interacting galaxies which examines their individual SFR properties.
There are a few studies which, while they contain no direct \hi~observations, interpret their results in the context of gas fraction. Studies of SFR triggering in high redshift galaxies (presumed to be at higher gas fractions) have found evidence for moderate SFR triggering \citep{Jogee2009, Robaina2009, Wong2011, Whitaker2014}, but {\textit{cf.}} \citet{Xu2012}, which found no evidence for SFR enhancement.
Historically, most direct studies of the gas content of interacting galaxies have been limited to detailed studies of one or two galaxies \citep[e.g.,][]{Smith1991, Smith1994a, Hibbard1994, Duc1997, Hibbard2001, Sengupta2013}.  

There are several blind \hi~surveys of the sky which could be used to investigate galaxy interactions, such as the \hi~Parkes All-Sky Survey \citep[HIPASS;][]{Barnes2001} and the Arecibo Legacy Fast ALFA (Arecibo L-band Survey Array) survey \citep[ALFALFA;][]{Giovanelli2005, Haynes2011}. ALFALFA has significant overlap with the Sloan Digital Sky Survey (SDSS) footprint, and so could be very useful.
A number of studies have made use of samples of galaxies cross matched between ALFALFA and the SDSS \citep[e.g.,][]{Catinella2012, Stierwalt2014, Fertig2015, Ellison2015}, though none of them have investigated the relationship between \delsfr~and gas fraction for individual galaxies. 
This is largely due to resolution limitations of single dish telescopes, which are unable to resolve individual galaxies in close pairs, where SFRs are most likely to be enhanced; gas fractions could therefore only be investigated as a pairwise property.

In order to reach the spatial resolutions required to investigate the gas content of galaxies at small separations, interferometric observations are required.  We therefore take new observations with the NRAO's Karl G. Jansky Very Large Array (VLA) to obtain \hi~masses for a sample of 34 galaxies in pairs.  Our 17 galaxy pairs (34 individual galaxies) from the SDSS Data Release 7 \citep[DR7;][]{Abazajian2009} have pre-existing SFRs and stellar masses. With the pre-existing data products from the SDSS, we calculate gas fractions and determine whether a strong observational correlation exists between gas fraction and SFR enhancement.

In Section \ref{sec:sample}, we present our sample selection criteria.  In Section \ref{sec:DR}, we present the data reduction steps for the acquired \hi~data.  In Section \ref{sec:products}, we present the reduced data products from the VLA, along with the correlation between SFR enhancement and the gas content of the galaxies.  In Section \ref{sec:simulations}, we compare our observational results with a small suite of simulations.  In Section \ref{sec:discussion}, we discuss the results, and present our conclusions in Section \ref{sec:conclusions}.
Throughout this work, we assume $\Omega_M = 0.3$, $\Omega_{\Lambda}=0.7$
and H$_0$ = 70 \kms\ Mpc$^{-1}$.

\section{Sample Selection}
\label{sec:sample}
Our master sample of galaxy pairs comes from the catalogue of \citet{Patton2011}.  This catalogue contains $>23,000$ galaxies in spectroscopically confirmed pairs from the SDSS Data Release 7 (DR7) within \rp~$< 80$ \kpc, and with relative velocity differences (\delv)  $< 10,000$ \kms.  In order to exclude galaxies which are likely to be projected pairs, previous works have traditionally imposed a limit on the \delv~of the pair somewhere between \delv~$<300$ \kms~\citep[e.g.,][]{Scudder2012b} and \delv~$ <500$ \kms~\citep[e.g.,][]{Ellison2008}.  For the current work, we impose a limit of \delv~$<300$ \kms~onto the galaxy pairs catalogue.

We require that all potential target galaxies have well determined stellar masses as determined by \citet{Mendel2014} and SFRs determined by \citet{Brinchmann2004} for the SDSS DR7. Since the goal of this work is to examine the dependence of the interaction-triggered SFR enhancement on the gas fraction of the galaxy (which is defined as the \hi~mass relative to the stellar mass), these quantities are critical for our analysis.  To ensure that the SFRs determined by \citet{Brinchmann2004} are uncontaminated by any flux from an AGN, we also require that at least one galaxy in the pair be classified as star forming on the \citet{Baldwin1981} diagram, using the classification scheme described by \citet{Kauffmann2003}.  
In order to be placed on the classification diagram, the S/N of the four required emission lines must be $>1.0$\footnote{Although the minimum S/N permitted is 1.0, all of the galaxies in the final sample have S/N $> 3.0$ in the four emission lines required for classification.}.   Only one galaxy in our  sample of 17 pairs is classified as an AGN, and is paired with a galaxy which is classified as star forming. We note that all of our galaxies are classified as star forming by the slightly more permissive \citet{Kewley2001} classification.  The remainder of the sample are star forming galaxies. {We note that this criterion requires that all galaxies be emission line galaxies.  While this is not an explicit cut on morphology, the majority of the sample will tend towards late-type morphologies.}

We also impose a lower limit on the redshift of our galaxy pairs of $z>0.01$, as below that threshold, the deblending algorithm of the SDSS is prone to shredding galaxies into multiple objects.  An upper redshift limit of $z<0.06$ was applied in order to prevent the exposure times for the VLA from becoming prohibitively large.

In order to assemble a sample of galaxy pairs which is most likely to be showing signs of triggered star formation, we use the results of \citet{Scudder2012b} to impose additional criteria on our sample. \citet{Scudder2012b} found that galaxies in pairs are most likely to be strongly enhanced when found at small \rp~from their companion (i.e., \rp~$< 30$ \kpc), and in nearly equal mass mergers ($0.33<M_{host}/M_{companion} < 3.0$).  We accordingly impose the criterion that the galaxy pairs must be at separations $<$ 30 \kpc.  In order to prevent the sample from becoming unnecessarily small, we opt to relax the definition of a `major merger' slightly and impose the criterion $0.25<M_{host}/M_{companion} < 4.0$\footnote{When this sample was defined, the stellar masses (and by extension, the mass ratios) were the same as those used in \citet{Scudder2012b}.  Since the acquisition of the data, the mass estimates have been refined by \citet{Mendel2014}.  In general, this does not appreciably change the stellar masses, but for two pairs in our sample, the stellar masses were changed by a sufficient amount to push their mass ratio past the 4:1 limit.  They have not been excluded from the sample.}.  This slightly more permissive definition has been used in the literature in the past as an alternate criterion to select nearly equal mass mergers \citep[e.g.,][]{Woods2007, Lotz2010}.
  
Our final criterion was to eliminate the galaxy pairs which were too closely positioned on the sky for the VLA's C configuration to be able to spatially resolve the two galaxies.  The synthesised beam width of C configuration is $\sim$ 14 arcsec, assuming uniform weighting.  We therefore required that the central locations of the two galaxies be separated by $> 35$ arcsec (or 2.5 times the synthesized beam width).  In combination with the criterion that the galaxies be at physical separations $<$ 30 \kpc, this biases our final sample towards the low redshift end of the permitted range.  

The full set of sample selection requirements is summarised in Table \ref{tab:requirements}, and results in a sample of 17 galaxy pairs, containing 34 individual galaxies.  The galaxies in our sample are presented in Table \ref{tab:vla:pairs}, which lists the SDSS Object ID (objid), the \rp~between pairs in \kpc, the velocity difference between the pair in \kms, the mass ratio between the two galaxies, and the stellar masses of each galaxy.

\begin{table}
\begin{center}
\begin{tabular}{|l|r|}
\hline
\delv & $<$ 300 \kms\\
Redshift & 0.01 $< z <$ 0.06\\
Mass ratio &0.25$< M_1/ M_2 <$4.0\\
\rp &$<$ 30 \kpc \\
Angular separation &$>35$ arcsec\\
\hline
\end{tabular}
\caption{\label{tab:requirements}Sample selection requirements for the galaxy pairs sample, in relative velocity difference (\delv), redshift (z), mass ratio, projected separation (\rp), and angular sky separation.}
\end{center}
\end{table}

\begin{table}
\begin{center}
\begin{tabular}{|c|c|c|c|c|}
\hline
SDSS ObjID & r$_p$ & $\Delta$v  & Mass ratio& log M$_*$  \\
&(\kpc) & (\kms) &&(M$_{\odot}$)\\
\hline
587726033341776175 & 21.51 &  36 & 3.66 & 9.65 \\
587726033341776191 & &  &  & 10.21 \\
\hline
587739609695453284 &22.27 & 65 & 2.50 &9.77\\
587739609695453281 &  &  & &10.17\\
\hline
587744873717563559 & 20.18 & 122 & 4.26 &9.14\\
587744873717563471 &  &  &  &8.51\\
\hline
587727179536859247 & 15.45 & 111 & 3.70 &8.90\\
587727179536859227 & &  & &8.33\\
\hline
587729160043757697 & 22.03 & 48 & 2.90 &8.61\\
587729160043757707 & &  & &9.08\\
\hline
587741489815027774 & 26.78 & 91 & 1.12 &9.83\\
587741489815028146 & &  & &9.78\\
\hline
587739303684866183 & 20.2 & 293 & 3.03 &8.61\\
587739303684866173 & &  & &9.09\\
\hline
587726033308680234 & 28.57 & 71 & 1.27 &9.65\\
587726033308680320 & &  & &9.75\\
\hline
588018056204780081 & 29.46 & 208 & 4.64 &9.81\\
588018056204780049 & &  & &10.48\\
\hline
587729158970867777 & 16.77 & 47 & 1.31 &9.87\\
587729158970867792 & &  & & 9.76\\
\hline
587733605328093368 & 27.7 & 130 & 1.15 &9.18\\
587733605328093256 & &  &  &9.12\\
\hline
588848899908370674 & 23.46 & 16 & 2.20 &9.19\\
588848899908370505 & &  &  &9.53\\
\hline
588017605758025795 & 15.2 & 49 & 1.09 & 8.46 \\
588017605758025732 & &  & & 8.50\\
\hline
588017702411763744 & 22.04 & 68 & 3.66 &9.64\\
588017702411763872 & &  &  &9.08\\
\hline
587727178473930875 & 27.87 & 73 & 1.24 &8.94\\
587727178473930886 & &  & &8.85\\
\hline
587742901789589569 & 19.55 & 118 & 3.59 &9.73\\
587742901789589575 & &  &  &9.17\\
\hline 
588023670245949622 & 19.39 & 53 & 1.63 &8.35\\
588023670245949625 & &  &  &8.56\\
\hline
\end{tabular}
\end{center}
\caption{SDSS objids, projected separations, velocity differences, mass ratios, and stellar masses for each of the 17 galaxy pairs in the final sample.}
\label{tab:vla:pairs}
\end{table}

\section{Data Acquisition}
\label{sec:DR}
Of the 17 galaxy pairs in our final sample, we were awarded VLA C configuration time for 12 galaxy pairs (24 galaxies) under Proposal ID 12A-061, with a mixture of Priority A, B, and C time.  
Director's Discretionary Time was successfully obtained during Cycle 13A to obtain VLA C configuration data for the remaining 5 galaxy pairs (Proposal 13A-537).  Of the awarded time, almost all of the scheduling blocks were observed; only one target (588023670245949622 \& 588023670245949625) was missing on-source time when the telescope ceased C configuration observations.
In this section, we briefly summarise the setup of these observations.

\subsection{Telescope setup}
In both proposals 12A-061 and 13A-537, we used the C configuration of the VLA, which has a maximum baseline between antennas of 3.4 km.  C configuration is one of the more compact of the VLA's configurations, allowing it to be more sensitive to large scale structure on the sky, relative to the more extended configurations of the array (A \& B).  However, this large-scale sensitivity comes at the expense of angular resolution. C configuration has a synthesised beamwidth of $14$ arcsec with a sensitivity to large scale features up to $970$ arcsec in size. At the typical redshift of our sample ($z\sim0.02$), this large angular scale sensitivity corresponds to a physical size of $\sim390$ \kpc. 
The large synthesised beam of the array in this configuration means that in general, the \hi~flux of a galaxy will be contained within a single beam, seen roughly as a point source by the array.  As the galaxies are well separated both spatially and in velocity, there should not be contamination from the companion galaxy. 

For the data taken in 12A-061, we used the dual polarisation mode of the WIDAR correlator in Open Shared Risk Observing (OSRO) program.  The data were taken between February 05, 2012, and March 18, 2012. 
The full spectral coverage was divided into 2 sets of 8 subbands, where each subband in one set was offset from the second set by half the width of the subband.  This resulted in high signal to noise coverage across the entire frequency range probed.  The correlator integration time was 5 seconds. A summary of the telescope setup is presented in Table \ref{table:settings}.

The data acquired as part of project 13A-537 are also in dual polarisation mode, but the Open Shared Risk Observing mode used in 12A-061 had been retired as the telescope exited commissioning.  The data acquired as a part of 13A-537 were observed between July 26, 2013 and September 02, 2013. Instead of using multiple subbands offset from each other by half the width of the subband, in 13A-537, a single 16 MHz wide subband was used, covering the entire frequency range.  The 13A-537 observations were carried out at the same frequency resolution and time resolution as the data from 12A-061.
A summary of the data acquisition is listed in Table \ref{tab:vla:sample}; this Table contains the SDSS objids, the redshifts of each galaxy, the $r$-band bulge to total ratio of the galaxy \citep{Simard2011}, the coordinates in Right Ascension and Declination, the proposal ID under which the observations were taken, the date of the observation, and the duration of the observations.

\begin{table}
\begin{center}
\begin{tabular}{|l|r|}
\hline
Receiver:& L Band 1000 - 2000 MHz\\
Rest Frequency &1420.40575  MHz\\
Sub-band Bandwidth &	4.0  MHz\\
No. of poln. products:&2.0  \\
No. of channels/poln product&	128  \\
Channel Width (Frequency)&	15.625  kHz\\
Channel Width (Velocity)& 	3.30 \kms @ 1420 MHz\\
\hline
\end{tabular}
\end{center}
\caption{Project 12A-061 telescope setup}
\label{table:settings}
\end{table}

\subsection{Data reduction}
The VLA data were reduced and calibrated using the Common Astronomy Software Applications \citep[{\sc CASA };][]{McMullin2007} and the Astronomical Image Processing System \citep[{\sc AIPS};][]{Greisen2003} packages of the National Radio Astronomy Observatory (NRAO). The imaging and deconvolution were performed in {\sc CASA} for all data sets.

Following standard procedures, the data were flagged for instrumental issues, radio frequency interference, and nonfunctional antennas, along with any other major problems noted in the observing logs. Bandpass calibration, complex gain calibration, and flux scaling were performed on each data set. 

The data were converted from the Fourier domain into the image plane using CASA's {\sc{clean}} task, and the line-free channels in the data cube were used to model and subtract off the continuum emission with {\sc{uvcontsub}}.  {\sc{clean}} was then used to create image cubes.  For those data sets in Table \ref{tab:vla:sample} with multiple observations, {\sc{clean}} was used at this stage to combine the UV (Fourier) data into a single image cube.  All images were created at a resolution of 4 arcsec to a pixel.

For each data cube, a spectrum was extracted for each galaxy in the pair.  The optimal spectral region was selected, using the RA and Declination from the SDSS as a starting point.  If necessary, the image cube was smoothed to coarser velocity resolution.  Once the optimal spatial box and velocity resolution were determined, the spectrum for that galaxy was extracted in units of flux density per optical \kms. 

\begin{table*}
\begin{center}
\begin{tabular}{|c|c|c|c|c|c|c|c|c|}
\hline
SDSS ObjID & Redshift ($z$)& B/T$_r$&RA (J2000) & Dec (J2000) & Proposal ID & Date Observed & Duration\\
\hline
587726033341776175 & 0.027775 & 0.0 &14:58:21.64 & +02:32:48.94 & 12A-061 &  Feb 06 2012 & 30 min \\
587726033341776191 & 0.027899 & 0.0 & 14:58:23.24 & +02:32:18.82 &  & & & \\
\hline
587739609695453284 & 0.027482 &0.0 & 11:18:11.21 & +31:59:16.32 &12A-061 &   Feb 22 2012 & 30 min \\
587739609695453281 & 0.02726  & 0.238 &11:18:10.97 & +31:58:35.93 &  & & &\\
\hline
587744873717563559 & 0.019347 & 0.0 &09:01:30.48 & +12:39:43.63 &12A-061  &  Feb 20 2012 & 1 h\\
587744873717563471 & 0.019761 & 0.223 &09:01:33.86 & +12:39:31.67 &  & & &\\
\hline
587727179536859247 & 0.018063 & 0.0 &02:06:21.33 & $-$08:52:19.26 &12A-061  & Mar 17 2012  & 1 h 30 min\\
587727179536859227 & 0.018439 & 0.071 &02:06:20.21 & $-$08:52:57.50 & & & &\\
\hline
587729160043757697 & 0.023771 & 0.997 & 13:32:02.01 & +05:32:12.73 &12A-061  &  Feb 07 2012 & 30 min\\
587729160043757707 & 0.023608 & 0.0 &13:32:04.62 & +05:32:37.14 & & & &\\
\hline
587741489815027774 & 0.03761 & 0.0 &08:05:53.02 & +14:00:06.34 &12A-061 & Feb 06 2012  & 30 min \\
587741489815028146 & 0.037924 & 0.039 &08:05:55.42 & +13:59:59.04 &\\
\hline
587739303684866183 & 0.017681 & 0.417 &13:12:07.51 & +34:16:13.78 &12A-061 & Feb 07 2012  & 30 min\\
587739303684866173 & 0.016686 & 0.18 &13:12:05.26 & +34:15:23.16 &\\
\hline
587726033308680234 & 0.031671 & 0.752 & 09:55:22.79 & +02:34:14.25 &12A-061  &  Feb 09 2012  & 30 min\\
587726033308680320 & 0.031426 & 0.0 &09:55:23.34 & +02:34:58.81 &\\
\hline
588018056204780081 & 0.03083 & 0.047 & 17:05:01.61 & +23:09:26.43 &12A-061& Feb 05 2012 & 30 min \\
588018056204780049 & 0.030114 & 0.072 &17:04:59.9 & +23:10:08.57 &\\
\hline
587729158970867777 & 0.023125 & 0.0 &13:39:27.31 & +04:30:38.03 &12A-061 &  Feb 07 2012 & 30 min\\
587729158970867792 & 0.022964 & 0.495 &13:39:29.71 & +04:30:40.95 &\\
\hline
587733605328093368 & 0.026354 & 0.045 &13:59:27.75 & +58:19:33.72 &12A-061 &  Feb 06 2012  & 30 min\\
587733605328093256 & 0.025908 & 0.192 &13:59:31.93 & +58:18:52.55 &\\
\hline
588848899908370674 & 0.025371 & 0.0 &11:20:57.36 & $-$00:05:49.48 &12A-061 & Feb 10 2012  & 30 min\\
588848899908370505 & 0.025317 & 0.0 &11:20:57.06 & $-$00:05:03.73 &\\
\hline
588017605758025795 & 0.019123 & 0.063 &12:18:38.72 & +45:46:28.13 & 13A-537 & Aug 6 2013 & 30 min\\
588017605758025732 & 0.019291 & 0.224 &12:18:39.14 & +45:47:06.90 &   & Aug 09 2013 & 30 min\\
 &&&&&& Aug 10 2013 & 30 min\\
\hline
588017702411763744 & 0.029248 & 0.154 &14:48:19.69 & +09:07:02.14 & 13A-537  &  Jul 26 2013 & 30 min\\
588017702411763872 & 0.029015 & 0.022 &14:48:17.59 & +09:07:23.48 & \\
\hline
587727178473930875 & 0.033197 & 0.411 &03:45:43.95 & $-$06:54:46.02 & 13A-537 &  Aug 18 2013 & 30 min\\
587727178473930886 & 0.032945 & 0.205 &03:45:46.7 & $-$06:54:56.24 & &  Aug 19 2013 & 30 min \\
 &&&&&&  Aug 27 2013 & 30 min\\
\hline
587742901789589569 & 0.025918 & 0.906 &12:39:35.85 & +16:35:16.14 & 13A-537 & Aug 13 2013 & 30 min \\
587742901789589575 & 0.026322 & 0.0 &12:39:38.35 & +16:35:06.34 &\\
\hline
588023670245949622 & 0.021097 & 0.0 &12:20:37.91 & +20:40:21.19 & 13A-537& Aug 10 2013 & 30 min \\
588023670245949625 & 0.021278 & 0.655 &12:20:39.01 & +20:39:38.69 &&   Aug 13 2013 & 30 min \\
 &&&&&&  Sep 02 2013  & 30 min\\
\hline
\end{tabular}
\end{center}
\caption{SDSS objids, redshifts, $r$-band bulge to total ratio, RA \& Declination, proposal ID, date observed, and the duration of the observation block, for all galaxy pairs in the sample (34 galaxies in 17 pairs).}
\label{tab:vla:sample}
\end{table*}

\section{Data products \& analysis}
In this section we describe the data products used for the analysis in the remainder of this paper.  This includes the calculation both of the \hi~masses and of the upper limits to the \hi~masses, along with a more detailed description of the data products from the SDSS for our sample of galaxies in pairs.
\label{sec:products}
\subsection{\hi~masses}
\label{sec:hi_mass}
For each galaxy in our sample, \hi~masses were calculated using the standard equation: 

\begin{equation}
M_{HI}= \left(2.36 \times 10^5\right) \times D^2 \int{S(v)~dv}
\label{eq:hi}
\end{equation}

$S(v)$ is the emission flux in Jy \kms.  The complete integral $\int{S(v)~dv}$ represents the integrated line flux over the width of the line.  $D$ is the distance to the object in Mpc.  For this work, we calculate  $D$ as $D=cz/H_0$, which is applicable for low redshift objects, such as those contained within our sample.  

\begin{figure*}
   \centering
   \begin{minipage}[c]{8cm}
   \includegraphics[height=210px]{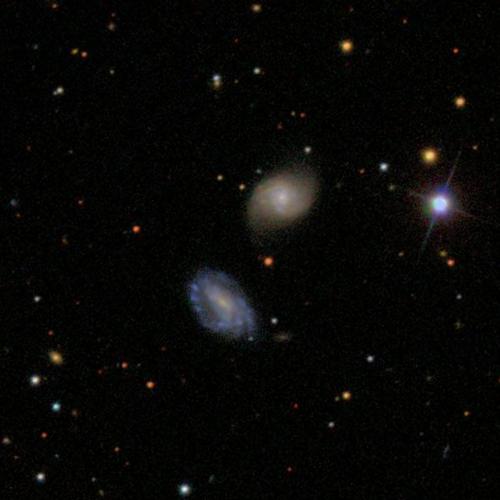} 
   \end{minipage}
   \begin{minipage}[c]{8cm}
   \includegraphics[height=215px]{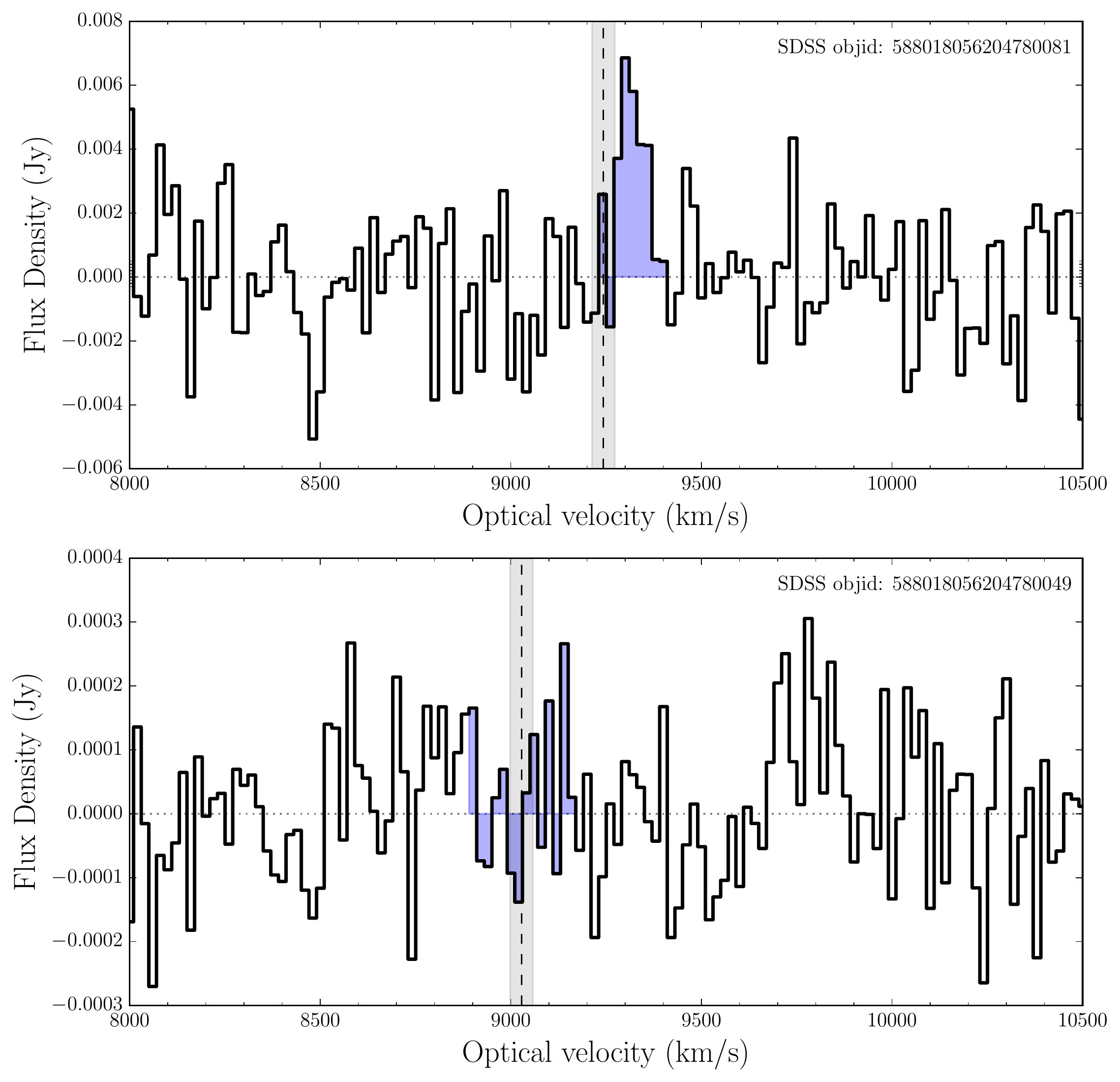}
    \end{minipage}
    \caption[~SDSS: 588018056204780081 \& 588018056204780049]{Left panel: SDSS thumbnail of galaxies 588018056204780081 (lower left) \& 588018056204780049 (upper right). The image is $250$ arcsec to a side.  Right panels: Flux density vs. optical velocity for galaxy pair 588018056204780081 (upper) \& 588018056204780049 (lower). The vertical dashed line indicates the systemic velocity of the galaxy, {with the gray shaded region indicating the uncertainty on the SDSS velocity ($\pm30$ \kms)} The horizontal dotted line indicates the line of zero flux. The blue shaded region indicates the region of the spectra used for signal to noise and gas mass calculations. The upper spectrum has a peak/RMS S/N of 3.19, while the lower has a S/N of 1.07.  The SDSS thumbnails and associated spectra for the remaining 16 galaxy pairs can be found in the Appendix. }
\label{fig:14352}
\end{figure*}

The significance of each \hi~mass was determined through a peak/RMS calculation.  The peak value of the emission line is divided by the RMS of the spectrum outside of the velocity range of the galaxy's emission.  In Figure \ref{fig:14352}, we show two of the 17 acquired spectra.  In the right hand side, we show the spectra for both galaxies in the pair, with the blue shaded region being the range of the spectrum used for peak/RMS calculations.  The vertical dashed line indicates the systemic velocity of the galaxy, with the grayscale shading indicating the uncertainty on the SDSS velocity\footnote{The uncertainty on the SDSS redshift is an average value for the Main Galaxy Sample, as quoted at the following URL: \url{http://classic.sdss.org/dr7/}} ($\pm 30$ \kms). The horizontal dotted line indicates the zero line.  On the left hand side of Figure \ref{fig:14352} we show the SDSS thumbnail image of the galaxy pair.  The remaining 16 galaxy pairs in the sample are shown in the Appendix (Figures \ref{fig:0875} through \ref{fig:402775}).  All calculated \hi~masses, with their peak/RMS signal to noise estimates, are found in Table \ref{tab:sn_sfr}.  

Of the 34 galaxies in the sample, 17 are detected at S/N $> 3.0$.  For galaxies whose spectrum has a marginal detection (peak/RMS S/N $ < 3.0$), an upper limit is determined using the bounds of the marginal spectral feature.  If no marginal detection is found, a velocity width of \delv$=300$ \kms, centred around the systemic velocity of the galaxy, is adopted.  We use upper limit values for the remainder of the analysis for all galaxies which were detected at S/N $<3$.  Upper limit calculations use the same velocity width used for the non-detection, and assumes $3\times$ the RMS value to be present across the entire velocity width.  This integrated flux value is then input into Equation \ref{eq:hi}, and the output is taken as a S/N = 3 upper limit.  These upper limits are included in Table \ref{tab:sn_sfr}.

\begin{table*}
\begin{center}
\begin{tabular}{|c|c|c|c|c|c|}
\hline
SDSS ObjID & \delsfr & log M$_*$ (M$_{\odot}$) & log M$_{HI}$ (M$_{\odot}$) & (M$_{HI}$/M$_{*}$) & S/N$_{(pk/RMS)}$ \\ \hline \hline
587726033341776175 &  0.25 & 9.65 & $<$ 9.32 & $<$0.47 &2.85 \\
587726033341776191 & $-$0.11 & 10.21 & 8.68 & 0.03 & 3.43  \\
\hline
587739609695453284 & 0.24  &9.77 & 9.19 & 0.26 & 4.08 \\
587739609695453281 & $-$0.42 &10.17 & 9.10 & 0.09 & 3.69\\
\hline
587744873717563559 & 0.001  & 9.14& 8.48 & 0.22 & 5.35 \\
587744873717563471 & 0.16 &8.51  &8.32 & 0.64 & 3.51 \\
\hline
587727179536859247 & $-$0.05  & 8.90 &9.48 & 3.82 & 11.40\\
587727179536859227 & 0.02 &8.33 & $<$ 8.78 & $<$ 2.81 & 2.02 \\
\hline
587729160043757697 &  0.60  & 8.61& 8.64 & 1.07 & 3.77 \\
587729160043757707 & 0.51  & 9.08& 8.70 & 0.42 & 3.16\\
\hline
587741489815027774 &  0.27  & 9.83 &  $<$ 9.92 & $<$ 1.23 & 2.70\\
587741489815028146 &0.54  & 9.78 &$<$ 9.84 & $<$ 1.15 & 2.23 \\
\hline
587739303684866183 &   $-$0.18 & 8.61 &$<$ 9.19 & $<$ 3.77 & 2.35 \\
587739303684866173 & $-$0.12 & 9.09 & 8.73 & 0.43 & 3.57\\
\hline
587726033308680234 &  0.32  & 9.65 & $<$ 9.94 & $<$ 1.97 & 1.97 \\
587726033308680320 &0.41  & 9.75 & $<$ 9.76 & $<$ 1.02 & 2.00\\
\hline
588018056204780081 &  0.18  &9.81 & 9.32 & 0.32 & 3.19 \\
588018056204780049 & 0.34  & 10.48& $<$ 8.64 & $<$ 0.01 & 2.17\\
\hline
587729158970867777 &0.02  & 9.87&  9.71 & 0.69 & 4.09 \\
587729158970867792 & $-$1.10  & 9.76 & 9.12 & 0.23 & 3.70 \\
\hline
587733605328093368 & $-$0.16 & 9.18 &  $<$ 8.84 & $<$ 0.46 & 2.26 \\
587733605328093256 & 0.28  &9.12  &9.66 & 3.53 & 4.04 \\
\hline 
588848899908370674 &  0.22  & 9.19 &  9.60 & 2.57& 4.54 \\
588848899908370505 & 0.15 &9.53  &  8.57 & 0.11& 3.26\\
\hline
588017605758025795 & $-$0.26  &8.46 &  $<$ 9.17 & $<$ 5.64 & 1.93  \\
588017605758025732 & 0.70   &8.50 &9.22 & 5.28 & 3.62 \\
\hline
588017702411763744 &  0.68   &9.64&  $<$ 10.25 & $<$ 4.07 & 2.67  \\
588017702411763872 & 0.082  &9.08&  $<$ 9.96 & $<$ 7.68 & 2.79 \\
\hline
587727178473930875 &  0.67   &8.94&  $<$ 10.02 & $<$ 11.90 & 2.62 \\
587727178473930886 & 0.14  &8.85& $<$ 9.50 & $<$ 4.46 & 2.24 \\
\hline
587742901789589569 &  0.54  &9.73&  9.54 & 0.66 & 3.20  \\
587742901789589575 & 0.16  & 9.17&  $<$ 9.99 & $<$ 6.53 & 2.70 \\
\hline 
588023670245949622 &  0.16  &8.35&   $<$ 9.01 & $<$ 4.51 & 1.61 \\
588023670245949625 & $-$0.31  &8.56& $<$ 9.39 & $<$ 6.68 & 2.73 \\
\hline
\end{tabular}
\caption{SDSS ObjID, \delsfr, stellar mass, \hi~gas mass, gas fraction, and the peak/RMS S/N for each of the galaxies in our sample. $3\sigma$ upper limits are included for all galaxies which were detected at S/N $<3$.  \label{tab:sn_sfr}}
\end{center}
\end{table*}

\subsubsection{Gas Fractions}
\label{sec:gasfraction}
Our estimation of the gas fraction uses the following expression, following previous observational works:

\begin{equation}
f_{gas}=\frac{M_{HI}}{M_*}
\label{eq:fgas}
\end{equation}
  
This is distinct from the calculation of most theoretical works, which prefer the form of:

\begin{equation}
f_{gas}=\frac{M_{gas}}{M_* + M_{gas}}
\label{eq:fgas_theory}
\end{equation}

The calculated gas fractions are presented in Table \ref{tab:sn_sfr}. 
We calculate gas fractions for all galaxies in the sample, including the upper limits, using the observational definition described in Equation \ref{eq:fgas}.  
  For consistency with the observational results presented in this paper, we have presented all theoretical values using the form $M_{gas}/M_*$, which is more directly comparable with the observational values, as the observations do not have total gas masses available.

In order to calculate the uncertainties on any individual \hi~detection in our sample, for those galaxies with peak/RMS S/N $>3.0$, we calculate a 1$\sigma$ error in a similar way to the calculation of the upper limits.  Instead of $3\times$ the RMS value of the spectrum outside the spectral line region, we take $1\times$ the RMS outside the emission region and calculate the equivalent \hi~mass associated with that flux density over the velocity width of the detected emission feature.

\subsection{SFR enhancements}
\label{sec:analysis}
As our sample of galaxy pairs was selected to be part of the SDSS DR7, each galaxy in the sample has a large amount of ancillary data available, which includes stellar masses \citep{Mendel2014} calculated from the updated photometry of \citet{Simard2011} and SFRs \citep{Brinchmann2004}.   
In order to correlate the gas fractions with the star formation enhancement in each galaxy, we now must define a metric which quantifies the amount of star formation triggered in each galaxy due to their interaction.

\subsubsection{SDSS Control Sample}
In order to estimate the change to the SFR of a galaxy in the pairs sample (described in Section \ref{sec:sample}) relative to what is `expected', we construct a robust control sample of galaxies which are not found in pairs.  These control galaxies are selected to be tightly matched to the pairs sample in stellar mass, redshift, and local density.
Local density is defined as in previous works \citep[e.g.,][]{Ellison2013} as:
\begin{equation}
\Sigma_n = \frac{n}{\pi d^2_n}
\end{equation}

$d_n$ is the distance to the $n^{th}$ nearest neighbour within a redshift slice of $\pm 1000$ \kms~around the redshift of the target galaxy; for this work, $n=5$.  $\Sigma_5$ is then normalised to the median value within a redshift range of $\pm 0.01$ to produce the normalised density ($\delta_5$).  $\delta_5$ is used to constrain the local density of the pairs and their control galaxies.

Matching a control sample in stellar mass, redshift, and local density has been found to eliminate the majority of sources of bias between the two samples \citep{Perez2009}.
Any remaining systematic issues within the data should be equally present in the pair sample and the control sample; any measurements of quantities which are relative between the two samples should not be affected.

To be considered in the matching process, each potential control galaxy must be classified as star forming on the BPT diagram according to the \citet{Kauffmann2003} classification scheme.  This criterion, as for the pairs sample, requires that the S/N of the necessary emission lines be $>1.0$, and excludes AGN and non-emission line galaxies.  The control galaxy must also not be included in the master pairs catalogue of \citet{Patton2011}, which removes all galaxies with a spectroscopic companion at \rp $< 80$ \kpc, and within \delv~$< 10,000$ \kms.  To further limit the spurious inclusion of galaxy pairs in the control sample which are missed due to spectroscopic incompleteness, we also require that the galaxy have a  merger vote fraction $f_m$ = 0, as determined by the GalaxyZoo participants \citep{Darg2010}.  \citet{Darg2010} use a cut in the vote fraction of $f_m$ = 0.4 (i.e., 40\% of Galaxy Zoo participants who inspected that particular galaxy classified it as an interacting system) to define a pairs sample.  In order to be conservative in our exclusion of galaxy pairs, we use the limit of $f_m = 0$ to ensure that interacting systems which were contentious or misclassified by the Galaxy Zoo participants are also excluded.

From this sample of potential control galaxies, we match each of the 34 galaxies in the VLA sample to a set of control galaxies which meet all of the criteria outlined above.  Following the work of \citet{Ellison2013}, control galaxies are matched to a paired galaxy if they fall within a redshift tolerance of $\Delta$z=0.005, a stellar mass tolerance $\Delta$log M$_*$=0.1 dex, and a local density tolerance of $\Delta\delta_5=0.1$, relative to that of the galaxy in the pair.  
Typically, each pair galaxy is matched to $>50$ control galaxies.  

Previous works \citep[e.g.,][]{Scudder2012, Scudder2012b, Ellison2013} have defined a quantity dubbed an `offset', denoted as \delsfr.  This offset value quantifies changes in SFR relative to the SFR -- stellar mass relation of galaxies.  
We follow the procedure of \citet{Ellison2013} in defining the offset value to be the difference between the SFR of the pair galaxy and the median SFR of the non-pair control galaxies.   
\begin{equation}
\Delta \mathrm{log(SFR)} = \mathrm{log}\left(\mathrm{SFR}_{pair}\right) - \mu_{1/2}\left(\mathrm{log}\left(\mathrm{SFR}_{controls}\right)\right)
\end{equation}

Positive values of \delsfr~indicate that the paired galaxy has a higher SFR than its matched controls, while negative values indicate that the galaxy has a lower SFR than its controls. We calculate offsets for the total, aperture corrected SFRs \citep{Brinchmann2004, Salim2007}, which provide the enhancement in the SFR across the entire disk.  

We obtain \delsfr~values for each of the galaxies in our pairs sample, which can then be individually plotted against the gas fractions calculated in Section \ref{sec:gasfraction}.  The median uncertainty in the SFR values among all star forming galaxies in the SDSS is 0.09 dex \citep{Brinchmann2004}; we adopt this value as the uncertainty in the \delsfr~across the pairs sample.

\subsection{Correlating \hi~mass with \delsfr}
\begin{figure}
   \centering
   \includegraphics[width=250px]{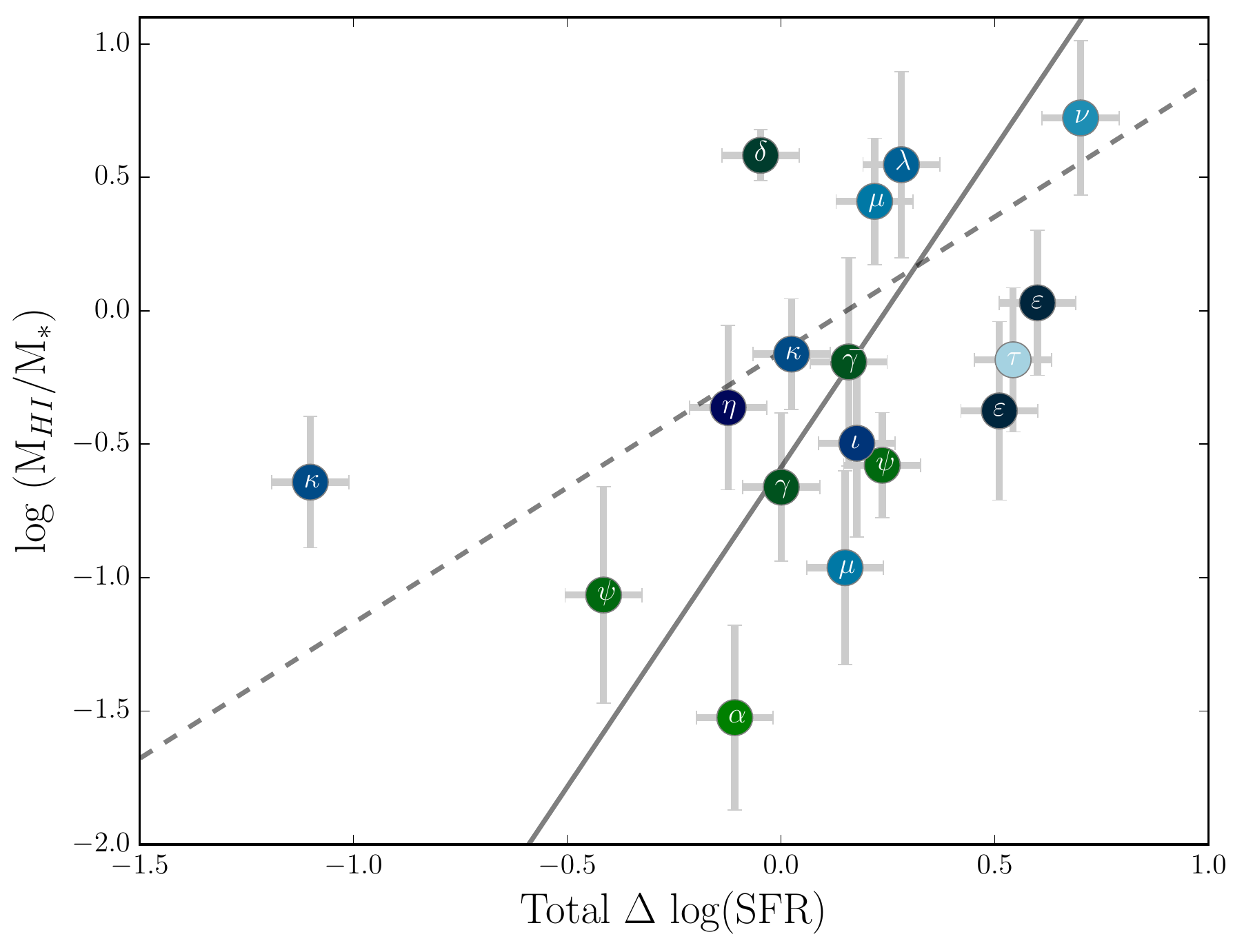}
   \caption[The log of the gas fraction vs. \delsfr, with least squares fit]{The log of the gas fraction vs.  \delsfr~for galaxies with S/N $>3.0$.  Galaxies are matched in colour and by greek letter if they are in the same pair.  The dashed grey line indicates the results of a least-squares fit to the observations, using the horizontal and vertical error bars as weights on the least squares fit, and is described by log$ \left(M_{HI}/M_{*}\right) $ = $ 1.02 \times $\delsfr$-0.16$. The solid dark grey line shows the results of the same fitting routine, but excludes the point plotted at \delsfr~$=-1.0$ (labeled with $\kappa$), and is described by log$ \left(M_{HI}/M_{*}\right)$=$2.39\times$\delsfr$-0.59$.}
   \label{fig:fgas_vs_tsfr_fit}
\end{figure}

 Figure \ref{fig:fgas_vs_tsfr_fit} shows the relation between gas fraction and \delsfr, for those galaxies in the sample which were detected at S/N~$>3.0$. The colours and greek letter of each data point matches the two galaxies in a pair, with each pair plotted as a unique colour. Error estimates in both gas fraction and \delsfr~are shown in grey. The full sample is presented in Figure \ref{fig:UL_gf_tsfr}. Galaxies with \hi~S/N~$<3.0$ (non-detections) are plotted with their $3\sigma$ upper limits, denoted by smaller triangles. In both figures, it appears that there is a rough trend of the highest gas fractions being found in galaxies which are most strongly starbursting.

To test the significance of this relationship between the total \delsfr~and the \hi~gas fraction, we make use of the Spearman Rank correlation test.  This correlation statistic produces a coefficient of 1.0 if the gas fraction of a galaxy always increases as it progresses to higher \delsfr, regardless of the shape of this correlation.  Other correlation metrics typically test for linear correlations, which may be less applicable in this instance.  A Spearman Rank correlation test of Figure \ref{fig:fgas_vs_tsfr_fit} produces a correlation coefficient of 0.5637. 
The Spearman Rank correlation test also outputs a null probability, which indicates the probability of no correlation, given the sample size. The probability of the data being uncorrelated produced by this test is 0.0184. This excludes the null hypothesis at slightly less than $2.5\sigma$.

In addition to the Spearman Rank correlation test, we also run a least-squares fitting routine on the data in order to acquire the an estimate of the slope of the data in order to quantify the correlation.  We use a line of the following form:

\begin{equation}
\mathrm{log} (M_{HI}/M_{*})=A*$\delsfr$+B
\end{equation}

\noindent The fitting procedure incorporates the errors in both gas fraction and in \delsfr~as weights on each point during the fitting.  The resulting line of best fit coefficients are A$=1.02 \pm 0.42$, B=$-0.16 \pm 0.17$, where the uncertainties are the standard errors calculated for each coefficient. This fit is plotted in Figure \ref{fig:fgas_vs_tsfr_fit} as the dashed grey line. However, we find that the yellow outlying point\footnote{Even though this galaxy has a very low overall \delsfr, it is still a star forming galaxy with significant (S/N $>3.0$) emission lines. This galaxy has low SFR in the region outside the SDSS fibre, so even though it has significant SFR within the fibre region, the overall SFR is shifted to a low value.} at \delsfr$=-1.0$ skews the fit towards a shallower slope and does not produce a representative fit to the majority of the points.  In order to more accurately fit the remaining points, we exclude this data point from the fitting procedure. (The exclusion of this data point still results in a Spearman rank null correlation exclusion of $>2\sigma$.)  The best fit relation to the data if the outlier is excluded has coefficients $A=2.39 \pm 0.82$ and $B=-0.59 \pm 0.26$.  This relation is plotted in Figure \ref{fig:fgas_vs_tsfr_fit} as the dark grey line. 

To test the sensitivity of the Spearman rank correlation test to the distribution of our data points given the small number statistics, we also perform a jackknife resampling procedure.  We remove one galaxy from the sample at a time, and re-perform the Spearman rank test on the 16 remaining galaxies, recording the null probability for each subsample.  This is repeated until each galaxy has been removed from the sample once.  We find that the probability of no correlation is remarkably robust to the removal of a single point; the median null probability of the subsamples is 0.0235, which is still well above 2$\sigma$\footnote{Two out of the 17 data points' removal results in a probability of null correlation that cannot be excluded at $>2\sigma$. The least constraining null probability is 0.0621, and the most 0.0030.}.  The probability of the data being uncorrelated, as calculated by the Spearman rank test, is therefore unlikely to be driven by single outliers.

\section{Simulations}
\label{sec:simulations}

To develop a more complete physical picture of what could drive the observed $\sim2.5\sigma$ correlation between increasing gas fraction and increasing \delsfr, we turn to simulations.  The existing theoretical work is not easily comparable to observations, as gas fractions are often measured in unobservable units, or compared at different time steps.  As a comparison to theoretical work may illuminate the physical mechanisms behind the observational results, we develop a small suite of simulations with the goal of comparing to observations.
\subsection{Simulation suite design}
\label{sec:simdesign}
Our suite is based on \citet{Torrey2012}, and a complete description of the model setup and parameters can be found in that work.  Briefly, the simulations of \citet{Torrey2012} use {\small{GADGET-2}} \citep{Springel2005a} to construct a set of N-body/smooth particle hydrodynamic models of binary galaxy mergers.  These mergers include radiative gas cooling~\citep{Katz1996}, star formation and associated feedback~\citep{Springel2003}, and chemical enrichment.  The SFR of the pair can be tracked either across the entire disk or within a smaller region at the centre of the disk; to compare to the current observations, we use the SFR across the entire disk.  The galaxies in the simulations are stable and do not develop a bar for at least 2 Gyr if they are evolved in isolation.  

In order to compare the results found here with the simulations, we construct a suite of binary mergers which vary only in gas fraction.  From \citet{Torrey2012}, we use orbit `e'.  The simulated galaxy models have initial stellar masses between $10.26<$ log M$_{\odot} < 10.52$. For each gas fraction, the galaxy model is merged with an identical copy of itself. We then track log(SFR) along with gas fraction for the duration of the simulation.
The suite uses initial gas fractions of $M_{gas}/M_*$ = 0.04, 0.09, 0.19, 0.32, 0.47, 0.67, 0.92, 1.27, and 1.78, for a total of 9 runs.  

\subsection{Results from the simulation suite}

\begin{figure*}
   \centering
   \includegraphics[width=480px]{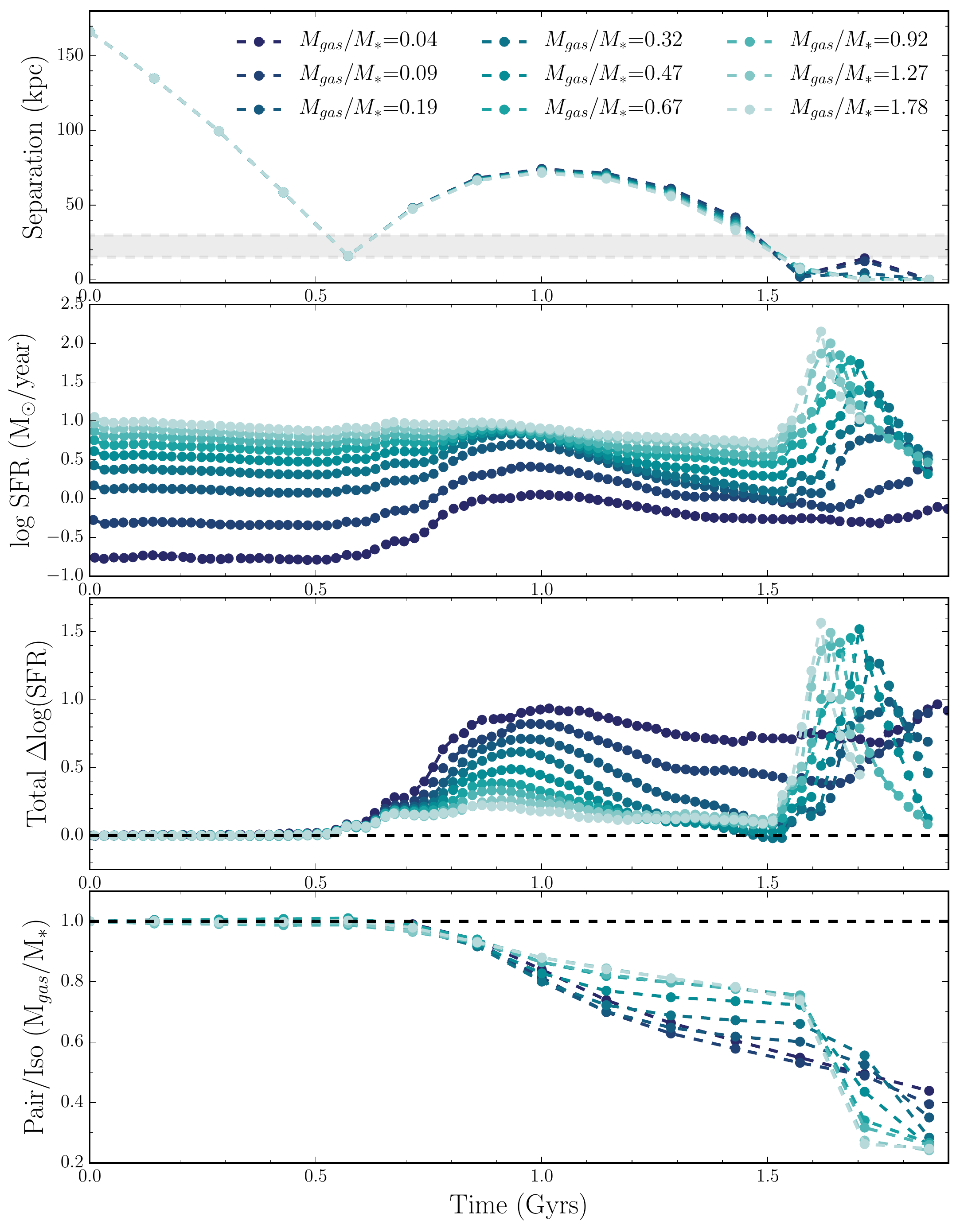}
   \caption[]{Results from our suite of 9 galaxy merger simulations.  In all panels, colours indicate the initial gas fraction, labeled in the top panel. In the top panel, the three dimensional separation between the two galaxies is plotted as a function of time. The galaxies all begin on the same orbit, so their tracks overlap until the final stages of the merger.  We truncate the data plotted here once the galaxies have a final separation of 0 \kpc. In the second panel, we plot the SFR for each galaxy as a function of time.  The third panel shows the \delsfr~from the simulations, normalised to the SFR of the isolated galaxy of the same initial gas fraction, as a function of time. In the lowest panel, we plot the ratio of the gas fraction of the interacting pair, relative to the gas fraction of the isolated galaxy at that time step. The horizontal dashed lines indicate where the interacting pairs and the isolated control are the same.}
   \label{fig:sims_v_time}
\end{figure*}

In the top panel of Figure \ref{fig:sims_v_time}, we show the time evolution of the physical (3D) separation between the two galaxies. 
Each colour indicates a unique initial gas fraction, labeled in the legend, progressing from navy for the most gas poor galaxy interaction ($M_{gas}/M_* =0.04$) to pale blue for the most gas rich ($M_{gas}/M_*=1.78$).  All of the galaxies have the same initial orbit; the only deviations are near coalescence, where slight discrepancies arise due to differences in dynamical friction.  

In the second panel, we show the log of the average SFR of the 2 interacting galaxies, as a function of time. For equal mass galaxies, the SFR response to the interaction should be approximately equal \citep[e.g.,][]{Torrey2012}. This panel indicates that higher gas fraction galaxies have systematically higher SFRs across the entire interaction (and prior to any encounter) than the low gas fraction galaxies.

In the third panel, we show the total \delsfr~for the galaxy in the interaction.  \delsfr~is calculated in a very similar manner to the observations.   
For each time step in the simulation\footnote{The second and third panels are both sampled at fixed time steps of 20 million years.}, we take the log(SFR) of the interacting galaxy, and subtract the log(SFR) of the same galaxy model in isolation at that same time step.  Prior to their first encounter, the two galaxies have the same SFR (\delsfr~= 0.0). The lowest gas fraction galaxies show rapid strong enhancement (relative to their isolated counterpart) after a close encounter, with the \delsfr~remaining high and nearly constant for the duration of the remaining merger.  However, the high gas fraction galaxies show a weak response to their initial pericentric passage, with strong \delsfr~only appearing in the final stages of the interaction.  

We lastly plot the evolution of the gas fraction in these galaxies, relative to their isolated counterpart.  For each time step, we divide the gas fraction of the pair by the gas fraction of the isolated galaxy.  By this definition, a value of 1.0 indicates that the two galaxies are consuming their gas at the same rate, which is the case before the first close passage.  As the pair galaxies consume their gas at a higher rate than they would in isolation in order to sustain a higher SFR, the ratio of the gas fractions declines. We note that for low gas fractions, the ratio drops more strongly than for high gas fractions; this is consistent with the third panel, which indicated that the low gas fraction galaxies had undergone a more significant change to their SFR, relative to their isolated counterpart, than the high gas fraction galaxies had.
For moderate gas fractions, the gas consumption is expected to be rather small; \citet{Fertig2015} find no evidence for large changes in gas fraction in their pairs sample, consistent with the results of prior simulations.

Figure \ref{fig:sims_v_time} indicates that the highest gas fraction galaxies show higher SFRs prior to an interaction than the lowest gas fraction galaxies.  In spite of the strong star formation prior to an interaction, high gas fraction galaxies are not strongly enhanced after a close passage.  By contrast, the low gas fraction galaxies show strong enhancements in their SFR, reaching enhancements of nearly a factor of 10 over their isolated counterparts shortly after a close encounter.   The $M_{gas}/M_* =0.04$ and $M_{gas}/M_* =0.09$ runs are dramatically elevated at all times after first pericentre.  
As the galaxies approach coalescence, the highest gas fraction galaxies now show a very strong response to coalescence, reaching SFR enhancements of a factor of 30 above their isolated controls (\delsfr~$=1.5$).  
The lowest gas fraction galaxies show a much weaker response to coalescence.  
Although all gas fractions show enhanced star formation after first pericentre, the relative strength of the low gas fraction starburst after first pericentre seems to indicate that high gas fractions are less efficient at forming new stars between first and second pericentre, but when the torques are maximal during coalescence, the larger gas volume may be able to create a larger burst of star formation. 

\begin{figure}
   \centering
   \includegraphics[width=250px]{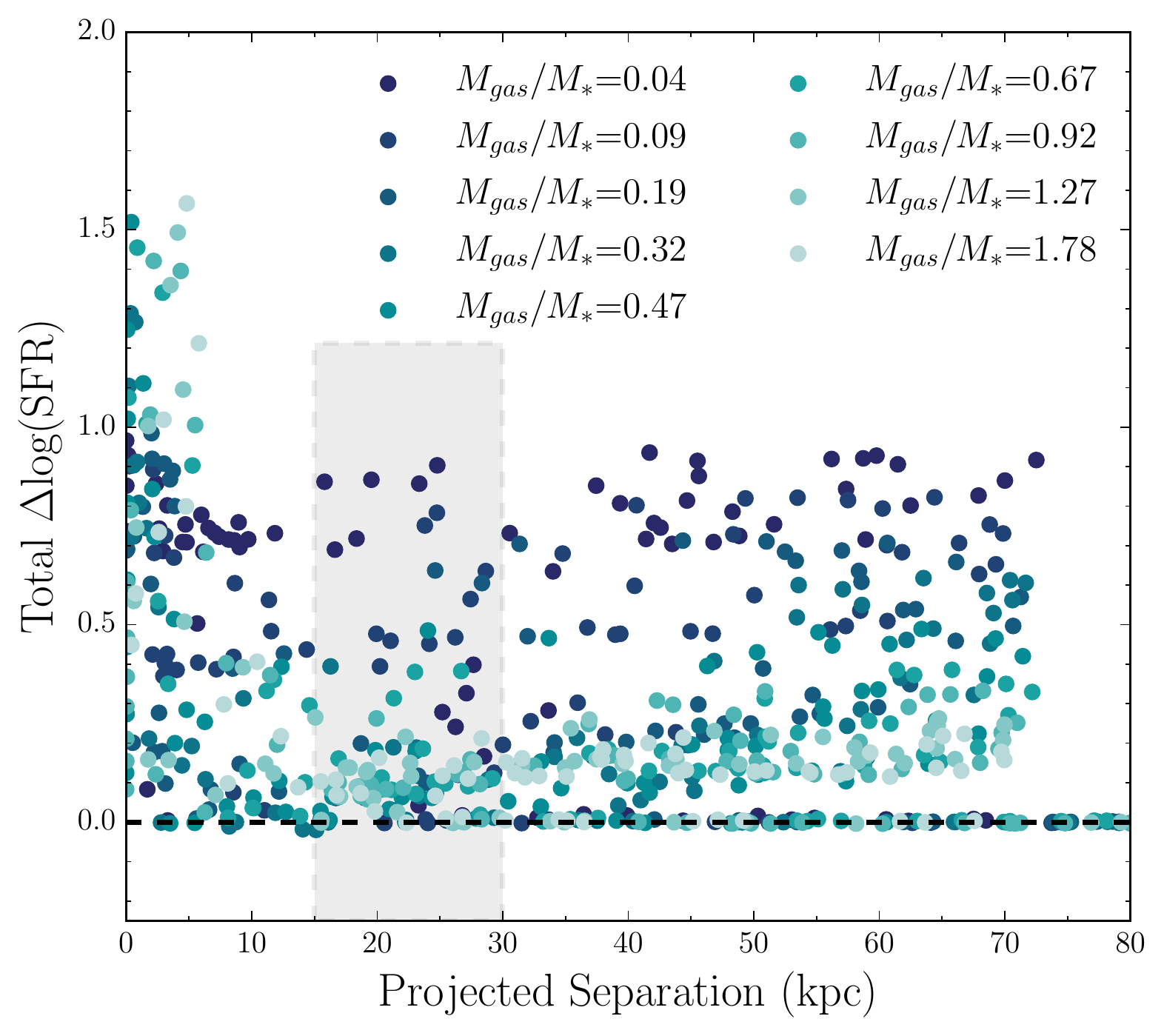}
   \caption[]{\delsfr~as a function of \rp.  As in Figure \ref{fig:sims_v_time}, colours indicate the initial gas fraction. Each data point from the third panel of Figure \ref{fig:sims_v_time} has had a random viewing angle applied to its separation to imitate the effects of observational projection effects, which tends to make galaxies appear closer in the sky than their true physical separation. The grey shaded region indicates the region spanned by the observational sample.}
   \label{fig:sims_sfr_rp}
\end{figure}

To compare more closely with the observations, we take two further steps with the simulations. First, we convert the real separations into projected separations, in order to mimic the effects of projection present in the observations, following the methodology of \citet{Scudder2012b} and \citet{Patton2013}. For each point in the simulations (corresponding to individual time steps), a 3 dimensional random viewing angle is applied to the separation data.  The 3-D angle is defined as $| \mathrm{cos} (\phi )|$, where $\phi=sin^{-1}(R)$ and $R$ is a different random value between 0 and 1 for each data point.  The cos($\phi$) term accounts for a 2 dimensional spin of the viewing angle, and the arcsin within the $\phi$ term accounts for the 3 dimensional distribution of angles across a sphere. This is repeated for each time step, which produces a new random viewing angle for each of the data points in the simulation. 
This conversion from real into projected separations preferentially scatters data points to smaller separations.
The projected separations are then plotted against their \delsfr~values in Figure \ref{fig:sims_sfr_rp}.    

\begin{figure}
   \centering
   \includegraphics[width=250px]{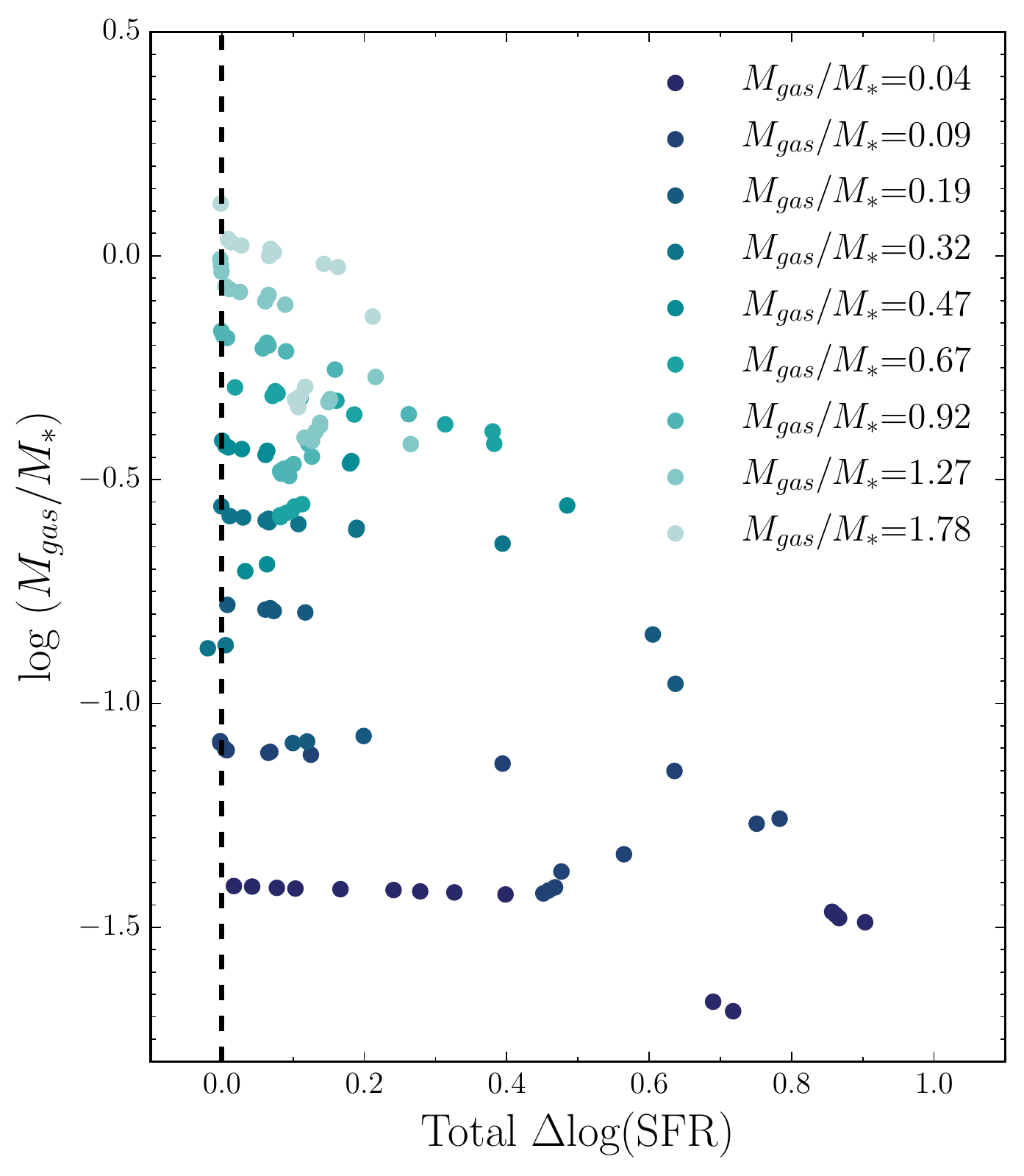}
   \caption[]{log(M$_{gas}$/M$_*$) as a function of \delsfr, for points which fall between 15 and 30 \kpc~after projection effects.  Colours indicate the initial gas fraction of the simulation, shown in the legend.  Galaxies with the highest instantaneous gas fractions are not typically found to have strong \delsfr~enhancements within this range of \rp.  However, the galaxies with low instantaneous gas fractions show a wide range of \delsfr, which include enhancements up to $\sim0.9$.}
   \label{fig:sims_fgas_sfr}
\end{figure}

The galaxy pairs in our VLA sample are explicitly limited to \rp~$<$ 30 \kpc.  However, the angular separation required for the VLA to spatially resolve the sample imposed an indirect lower limit on \rp,  such that all galaxies in the sample fall in a relatively narrow range of \rp: $15 <$ \rp~$<30$ \kpc.  To mimic this constraint in the simulations, we select out the subsample of the simulation data which falls within projected separations of between 15 and 30 \kpc, indicated in Figure \ref{fig:sims_sfr_rp} as the grayscale region.
As the observations are also limited to instantaneous gas fractions, we plot in Figure \ref{fig:sims_fgas_sfr} the instantaneous gas fraction of the galaxy versus \delsfr, for those time steps which placed the data points within $15<$ \rp~$<30$ \kpc.

Figure \ref{fig:sims_fgas_sfr} illustrates \delsfr~as a function of log($M_{gas}/M_*$).  As this region of \rp~contains both galaxies which have not yet had a close pass, those just after their close pass, and galaxies returning towards coalescence, there is a spread in \delsfr~values for any given value of $M_{gas}/M_*$. However, it is clear that those galaxies in the lowest $M_{gas}/M_*$ regime are also those galaxies which permit the strongest \delsfr~enhancements.
 In comparing Figure \ref{fig:sims_fgas_sfr} to the results found based upon the observational data in Figure \ref{fig:fgas_vs_tsfr_fit}, the correlations of the simulations and the observations appear to be in contrary directions.  We discuss this apparent contradiction further in the Discussion section.

\section{Discussion}	
\label{sec:discussion}

The current observational results suggest a positive correlation between the \hi~content of a galaxy and the strength of triggered star formation within that galaxy.  Spearman rank correlation tests indicate that the data exclude the null hypothesis at $\sim2.5\sigma$ significance, and a jackknife resampling of the data indicates that the correlation is not extremely sensitive to individual galaxies in the sample.
In contrast, simulations of binary mergers with varying gas fractions indicate that prior to coalescence, we would expect an anti-correlation between gas fraction and SFR enhancement, relative to an identical galaxy in isolation.
We now wish to investigate what may lie at the root of this difference.

\subsection{Reconciling the different physical pictures}

There is a critical difference between the simulations and the current observational data; for the simulations, the control is an exact match in all parameters to the interacting galaxy prior to its interaction.  The observational data is very tightly matched in stellar mass, redshift, and local environment, but, importantly, the gas fractions of the SDSS control galaxies are unconstrained.  

\begin{figure}
   \centering
   \includegraphics[width=250px]{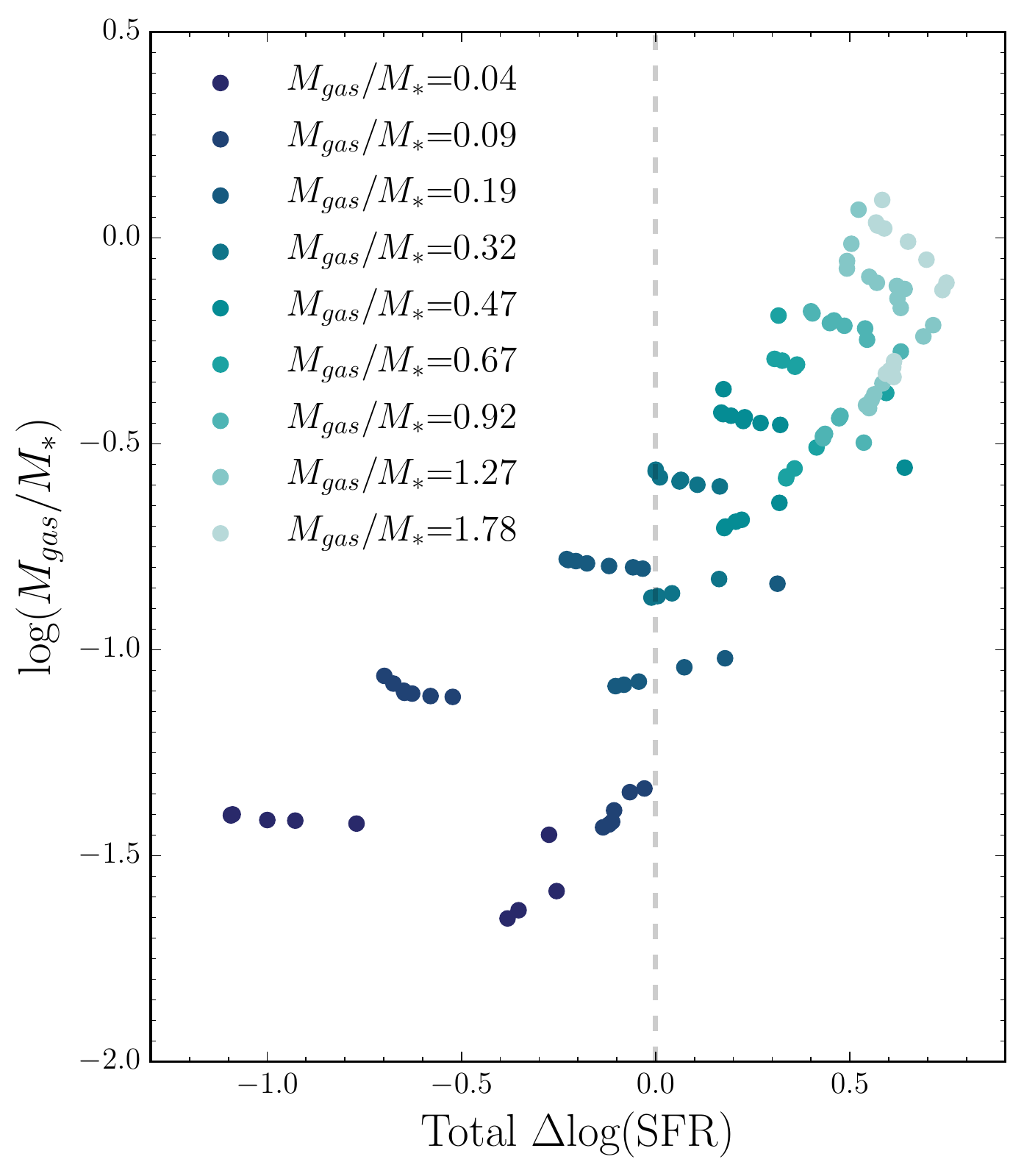}
   \caption[]{A counterpart to Figure \ref{fig:sims_fgas_sfr}, with \delsfr~values calculated relative to a fixed gas fraction isolated galaxy (\fgas$=0.32$). There now appears to be a strong correlation between the initial gas fraction and the level of SFR enhancement.}
   \label{fig:sfr_v_fgas_fixctl}
\end{figure}

We test the impact of leaving the gas fraction unconstrained in the observational controls using the simulation data.  We choose a single simulated galaxy (in isolation) to operate as the `average' control galaxy for all of the interacting galaxies. 
We then recalculate the \delsfr~using this fixed control galaxy for all simulated encounters.  This means that every interacting galaxy's SFR is being compared to the same isolated galaxy, which has the same stellar mass but a different gas fraction.
For this recalculation of \delsfr, we select the \fgas~$=0.32$ run (an intermediate gas fraction) as our `average' control.  In all other aspects, the calculation proceeds as before.
We regenerate Figure \ref{fig:sims_fgas_sfr} with the recalculated \delsfr~values in Figure \ref{fig:sfr_v_fgas_fixctl}.  Figure \ref{fig:sfr_v_fgas_fixctl} shows the instantaneous gas fraction as a function of \delsfr~for galaxies within $15<$\rp$< 30$ \kpc, now using the fixed control galaxy \fgas~$=0.32$ for the calculation of \delsfr. 

A comparison of Figure \ref{fig:sims_fgas_sfr} and Figure \ref{fig:sfr_v_fgas_fixctl} demonstrates that the lack of a gas fraction match in the control sample drastically changes the apparent relationship between gas fraction and \delsfr~from an anti-correlation to a correlation.  By using a galaxy with an average gas fraction as the simulation's control, the trend between gas fraction and \delsfr~in the simulations is now in broad agreement with the observations. As demonstrated in the second panel of Figure \ref{fig:sims_v_time}, there is a strong pre-interaction correlation between gas fraction and SFR, which can explain this change.
Although galaxies with low initial gas fractions exhibit the highest interaction-triggered enhancements, they begin an interaction with low absolute SFRs. These initially low absolute SFRs translate to strongly negative \delsfr s when compared with an average gas fraction control.  For these low gas fraction systems, even the strong increase in their SFRs during the interaction is still not enough of a boost to increase their SFRs above the non-interacting SFR of an average gas fraction control.  
Conversely, high gas fraction galaxies have a pre-interaction SFR that is already enhanced relative to the average gas fraction control.  Although their interaction-induced enhancement is modest, they have such strong SFR at the outset that they still exhibit the highest \delsfr s.
For galaxies without a control sample explicitly matched in gas fraction, the correlation between the gas fraction of a galaxy and its SFR, independent of the interaction environment, is clearly much stronger than the correlation between the gas fraction and the triggered SFR due to an interaction.

In the ideal case, this test would lead us to construct a more suitable control sample for our observational sample in order to remove this source of bias.  
However, we would require a large sample with appropriately deep \hi~observations in order to construct a control sample matched in gas fraction such a sample does not currently exist.
ALFALFA.40 \citep{Haynes2011}  is the largest sample of galaxies with cross-matches to the SDSS, which would normally produce the best number statistics for the construction of a control sample.  However, our VLA sample is significantly deeper than the ALFALFA.40  sample at a given stellar mass.  This difference in survey depth means that we would only be able to find gas fraction matches to the highest gas fraction galaxies at fixed stellar mass, and these are the galaxies that the simulations indicate are least likely to show a strong enhancement.  Without the ability to match our entire VLA sample, the ALFALFA sample of galaxies is not useful as a source of control galaxies.

We note that the simulation suite presented here provides a wide range of gas fractions at fixed stellar mass, and therefore should serve as an upper limit to the strength of correlation driven by poor matching in gas fraction.  Even though log(\fgas) spans more than an order of magnitude in the simulations, a similar range of \hi~gas fractions at fixed stellar mass has been observed in large \hi~surveys \citep[e.g., ][]{Catinella2012, Huang2012a}, indicating that simple stellar mass matching is unlikely to indirectly control for the gas fraction of the galaxy, and this source of bias is almost certainly present in the observational sample. 

\subsection{The Impact of Molecular Gas}
As described in Section \ref{sec:hi_mass}, our adopted definition of gas fraction in this work does not include any contribution from the molecular gas component of a galaxy. However, the impact of the exclusion of molecular gas is not well understood. 
\citet{Saintonge2011} finds that for non-interacting galaxies, the molecular gas mass of a galaxy is about 30\% of the neutral gas mass on average, with a great deal of scatter; the two quantities are, at best, only weakly correlated.
The behaviour of the molecular gas in interacting galaxies is still very unclear. It has been suggested that in galaxy mergers, the relationship between SFR and the molecular gas content is systematically changed such that interacting galaxies have significantly faster depletion times; i.e., they turn more gas into stars per unit time, when compared to their isolated counterparts \citep{Daddi2010, Genzel2010, Saintonge2012}, with \citet{Stark2013} finding that recently enhanced SFR (such as is present in interactions) correlates with a larger H$_2$/\hi~content. However, \citet{Zasov1994} finds that interacting galaxies have only a small component of molecular gas, and \citet{Casasola2004} found that galaxies in pairs have normal \hi~content but higher than average molecular gas (resulting in a higher than average total gas mass for interacting galaxies). Until a direct test of the dependence of SFR on the H$_2$ content of a sample of galaxy pairs is undertaken, it will remain unclear how the molecular gas will correlate with the SFR enhancement of a galaxy. 

The measurements from the simulations presented here (to which we are comparing the observations) are akin to the the total gas mass of the galaxy, with some fraction of the total gas mass divided between \hi~and H$_2$.  While it would be interesting to examine the impact of molecular gas in simulated interactions, this is beyond the scope of the current work.
\section{Conclusions}
\label{sec:conclusions}

In this work, we have attempted to determine if the gas content of a galaxy plays a dominant role in influencing the strength of an interaction-triggered starburst. 
To this end, we present new observational data from the VLA, which has allowed us to estimate the \hi~content of a sample of 34 galaxies in 17 galaxy pairs.  Of the 34 galaxies, 17 are detected with S/N $> 3.0$. The detected \hi~mass contained within these galaxies was converted into a gas fraction, using the stellar masses of \citet{Mendel2014}.  

We compare the gas fraction of these galaxies with their SFR enhancements (\delsfr), which are determined relative to a control sample which is tightly matched in stellar mass, redshift, and local density.  
We find that the \hi~gas fraction of the VLA galaxies is correlated at $\sim2.5\sigma$ with the total (i.e., across the entire galaxy) star formation enhancement of that galaxy.  
We then compare the observational results with a suite of simulations in order to help develop an interpretive framework for the observations.

The simulation suite consists of a set of 9 simulated galaxy mergers which vary only in gas fraction. The simulations suggest that the lowest gas fraction galaxies should show the highest enhancements to their SFR after their first encounter with their companion.  High gas fraction galaxies appear the least affected by their close pass, relative to an identical, non-interacting galaxy, implying an anti-correlation between \delsfr~and gas fraction.

The simulations also indicate that high gas fraction galaxies have systematically higher pre-interaction SFRs when compared to low gas fraction galaxies.  We determine that if the gas fractions of the control galaxies are not part of the matching procedure (as is the case for our observational sample), an apparent correlation between \delsfr~and gas fraction can be produced, as the correlation between higher base SFR and higher gas fraction is much stronger than the anti-correlation of \delsfr~and gas fraction due to the interaction itself. 
If the simulations are handled in a manner consistent with the observations, we find the same qualitative correlation between gas fraction and SFR enhancement.

\section*{Acknowledgments}
We thank the anonymous referee for constructive comments which improved the clarity of this paper. 

JMS wishes to extend particular thanks to Miriam Krauss at the NRAO Helpdesk for her assistance with the early stages of the data reduction process, and to the NRAO Domenici Science Operations Center in Socorro for their hospitality while the majority of this data reduction was completed.  We also thank Paola Di Matteo, Chris Pritchet, and Ruben Sanchez-Janssen for helpful comments on an early version of this work.

SLE and DRP acknowledge the receipt of an NSERC Discovery grant which funded this research. DF and JLR acknowledge NSF grant AST-000167932 and George Mason University Presidential Fellowship for support of this work.

The National Radio Astronomy Observatory is a facility of the National Science Foundation operated under cooperative agreement by Associated Universities, Inc.

We are grateful to the MPA/JHU group for access to their data products and catalogues (maintained by Jarle Brinchmann at  \url{http://www.mpa-garching.mpg.de/SDSS/}).  

Funding for the SDSS and SDSS-II has been provided by the Alfred P. Sloan Foundation, the Participating Institutions, the National Science Foundation, the U.S. Department of Energy, the National Aeronautics and Space Administration, the Japanese Monbukagakusho, the Max Planck Society, and the Higher Education Funding Council for England. The SDSS Web Site is \url{http://www.sdss.org/}.
The SDSS is managed by the Astrophysical Research Consortium for the Participating Institutions. The Participating Institutions are the American Museum of Natural History, Astrophysical Institute Potsdam, University of Basel, University of Cambridge, Case Western Reserve University, University of Chicago, Drexel University, Fermilab, the Institute for Advanced Study, the Japan Participation Group, Johns Hopkins University, the Joint Institute for Nuclear Astrophysics, the Kavli Institute for Particle Astrophysics and Cosmology, the Korean Scientist Group, the Chinese Academy of Sciences (LAMOST), Los Alamos National Laboratory, the Max-Planck-Institute for Astronomy (MPIA), the Max-Planck-Institute for Astrophysics (MPA), New Mexico State University, Ohio State University, University of Pittsburgh, University of Portsmouth, Princeton University, the United States Naval Observatory, and the University of Washington.

\bibliographystyle{apj}
\bibliography{masterfile_bibdesk}

\begin{thebibliography}{79}
\expandafter\ifx\csname natexlab\endcsname\relax\def\natexlab#1{#1}\fi

\bibitem[{{Abazajian} {et~al.}(2009){Abazajian}, {Adelman-McCarthy},
  {Ag{\"u}eros}, {Allam}, {Allende Prieto}, {An}, {Anderson}, {Anderson},
  {Annis}, {Bahcall}, \& et~al.}]{Abazajian2009}
{Abazajian}, K.~N., {et~al.} 2009, ApJS, 182, 543

\bibitem[{{Alonso} {et~al.}(2006){Alonso}, {Lambas}, {Tissera}, \&
  {Coldwell}}]{Alonso2006}
{Alonso}, M.~S., {Lambas}, D.~G., {Tissera}, P., \& {Coldwell}, G. 2006, MNRAS,
  367, 1029

\bibitem[{{Alonso} {et~al.}(2004){Alonso}, {Tissera}, {Coldwell}, \&
  {Lambas}}]{Alonso2004}
{Alonso}, M.~S., {Tissera}, P.~B., {Coldwell}, G., \& {Lambas}, D.~G. 2004,
  MNRAS, 352, 1081

\bibitem[{{Baldry} {et~al.}(2006){Baldry}, {Balogh}, {Bower}, {Glazebrook},
  {Nichol}, {Bamford}, \& {Budavari}}]{Baldry2006}
{Baldry}, I.~K., {Balogh}, M.~L., {Bower}, R.~G., {Glazebrook}, K., {Nichol},
  R.~C., {Bamford}, S.~P., \& {Budavari}, T. 2006, MNRAS, 373, 469

\bibitem[{Baldwin {et~al.}(1981)Baldwin, Phillips, \& Terlevich}]{Baldwin1981}
Baldwin, J.~A., Phillips, M.~M., \& Terlevich, R. 1981, PASP, 93, 5

\bibitem[{{Barnes} {et~al.}(2001){Barnes}, {Staveley-Smith}, {de Blok},
  {Oosterloo}, {Stewart}, {Wright}, {Banks}, {Bhathal}, {Boyce}, {Calabretta},
  {Disney}, {Drinkwater}, {Ekers}, {Freeman}, {Gibson}, {Green}, {Haynes}, {te
  Lintel Hekkert}, {Henning}, {Jerjen}, {Juraszek}, {Kesteven}, {Kilborn},
  {Knezek}, {Koribalski}, {Kraan-Korteweg}, {Malin}, {Marquarding}, {Minchin},
  {Mould}, {Price}, {Putman}, {Ryder}, {Sadler}, {Schr{\"o}der}, {Stootman},
  {Webster}, {Wilson}, \& {Ye}}]{Barnes2001}
{Barnes}, D.~G., {et~al.} 2001, MNRAS, 322, 486

\bibitem[{{Barnes} \& {Hernquist}(1996)}]{Barnes1996}
{Barnes}, J.~E., \& {Hernquist}, L. 1996, ApJ, 471, 115

\bibitem[{{Bournaud} {et~al.}(2011){Bournaud}, {Chapon}, {Teyssier}, {Powell},
  {Elmegreen}, {Elmegreen}, {Duc}, {Contini}, {Epinat}, \&
  {Shapiro}}]{Bournaud2011}
{Bournaud}, F., {et~al.} 2011, ApJ, 730, 4

\bibitem[{{Brinchmann} {et~al.}(2004){Brinchmann}, {Charlot}, {White},
  {Tremonti}, {Kauffmann}, {Heckman}, \& {Brinkmann}}]{Brinchmann2004}
{Brinchmann}, J., {Charlot}, S., {White}, S.~D.~M., {Tremonti}, C.,
  {Kauffmann}, G., {Heckman}, T., \& {Brinkmann}, J. 2004, MNRAS, 351, 1151

\bibitem[{{Casasola} {et~al.}(2004){Casasola}, {Bettoni}, \&
  {Galletta}}]{Casasola2004}
{Casasola}, V., {Bettoni}, D., \& {Galletta}, G. 2004, A\&A, 422, 941

\bibitem[{{Catinella} {et~al.}(2012){Catinella}, {Schiminovich}, {Kauffmann},
  {Fabello}, {Hummels}, {Lemonias}, {Moran}, {Wu}, {Cooper}, \&
  {Wang}}]{Catinella2012}
{Catinella}, B., {et~al.} 2012, A\&A, 544, A65

\bibitem[{{Cox} {et~al.}(2006){Cox}, {Jonsson}, {Primack}, \&
  {Somerville}}]{Cox2006}
{Cox}, T.~J., {Jonsson}, P., {Primack}, J.~R., \& {Somerville}, R.~S. 2006,
  MNRAS, 373, 1013

\bibitem[{{Cox} {et~al.}(2008){Cox}, {Jonsson}, {Somerville}, {Primack}, \&
  {Dekel}}]{Cox2008}
{Cox}, T.~J., {Jonsson}, P., {Somerville}, R.~S., {Primack}, J.~R., \& {Dekel},
  A. 2008, MNRAS, 384, 386

\bibitem[{{Daddi} {et~al.}(2010){Daddi}, {Elbaz}, {Walter}, {Bournaud},
  {Salmi}, {Carilli}, {Dannerbauer}, {Dickinson}, {Monaco}, \&
  {Riechers}}]{Daddi2010}
{Daddi}, E., {et~al.} 2010, ApJL, 714, L118

\bibitem[{{Darg} {et~al.}(2010){Darg}, {Kaviraj}, {Lintott}, {Schawinski},
  {Sarzi}, {Bamford}, {Silk}, {Andreescu}, {Murray}, {Nichol}, {Raddick},
  {Slosar}, {Szalay}, {Thomas}, \& {Vandenberg}}]{Darg2010}
{Darg}, D.~W., {et~al.} 2010, MNRAS, 401, 1552

\bibitem[{{Di Matteo} {et~al.}(2007){Di Matteo}, {Combes}, {Melchior}, \&
  {Semelin}}]{diMatteo2007}
{Di Matteo}, P., {Combes}, F., {Melchior}, A.-L., \& {Semelin}, B. 2007, A\&A,
  468, 61

\bibitem[{{Duc} {et~al.}(1997){Duc}, {Brinks}, {Wink}, \& {Mirabel}}]{Duc1997}
{Duc}, P.-A., {Brinks}, E., {Wink}, J.~E., \& {Mirabel}, I.~F. 1997, A\&A, 326,
  537

\bibitem[{{Ellison} {et~al.}(2015){Ellison}, {Fertig}, {Rosenberg}, {Nair},
  {Simard}, {Torrey}, \& {Patton}}]{Ellison2015}
{Ellison}, S.~L., {Fertig}, D., {Rosenberg}, J.~L., {Nair}, P., {Simard}, L.,
  {Torrey}, P., \& {Patton}, D.~R. 2015, MNRAS, 448, 221

\bibitem[{{Ellison} {et~al.}(2013){Ellison}, {Mendel}, {Scudder}, {Patton}, \&
  {Palmer}}]{Ellison2013}
{Ellison}, S.~L., {Mendel}, J.~T., {Scudder}, J.~M., {Patton}, D.~R., \&
  {Palmer}, M.~J.~D. 2013, MNRAS, 430, 3128

\bibitem[{{Ellison} {et~al.}(2011){Ellison}, {Patton}, {Mendel}, \&
  {Scudder}}]{Ellison2011b}
{Ellison}, S.~L., {Patton}, D.~R., {Mendel}, J.~T., \& {Scudder}, J.~M. 2011,
  MNRAS, 418, 2043

\bibitem[{{Ellison} {et~al.}(2008){Ellison}, {Patton}, {Simard}, \&
  {McConnachie}}]{Ellison2008}
{Ellison}, S.~L., {Patton}, D.~R., {Simard}, L., \& {McConnachie}, A.~W. 2008,
  AJ, 135, 1877

\bibitem[{{Ellison} {et~al.}(2010){Ellison}, {Patton}, {Simard}, {McConnachie},
  {Baldry}, \& {Mendel}}]{Ellison2010}
{Ellison}, S.~L., {Patton}, D.~R., {Simard}, L., {McConnachie}, A.~W.,
  {Baldry}, I.~K., \& {Mendel}, J.~T. 2010, MNRAS, 407, 1514

\bibitem[{Fertig {et~al.}(2015)Fertig, Rosenberg, Patton, Ellison, Torrey, \&
  Scudder}]{Fertig2015}
Fertig, D.~J., Rosenberg, J.~L., Patton, D.~R., Ellison, S.~L., Torrey, P., \&
  Scudder, J.~M. 2015, MNRAS, submitted

\bibitem[{{Genzel} {et~al.}(2010){Genzel}, {Tacconi}, {Gracia-Carpio},
  {Sternberg}, {Cooper}, {Shapiro}, {Bolatto}, {Bouch{\'e}}, {Bournaud},
  {Burkert}, {Combes}, {Comerford}, {Cox}, {Davis}, {Schreiber},
  {Garcia-Burillo}, {Lutz}, {Naab}, {Neri}, {Omont}, {Shapley}, \&
  {Weiner}}]{Genzel2010}
{Genzel}, R., {et~al.} 2010, MNRAS, 407, 2091

\bibitem[{{Giovanelli} {et~al.}(2005){Giovanelli}, {Haynes}, {Kent},
  {Perillat}, {Saintonge}, {Brosch}, {Catinella}, {Hoffman}, {Stierwalt},
  {Spekkens}, {Lerner}, {Masters}, {Momjian}, {Rosenberg}, {Springob},
  {Boselli}, {Charmandaris}, {Darling}, {Davies}, {Garcia Lambas}, {Gavazzi},
  {Giovanardi}, {Hardy}, {Hunt}, {Iovino}, {Karachentsev}, {Karachentseva},
  {Koopmann}, {Marinoni}, {Minchin}, {Muller}, {Putman}, {Pantoja}, {Salzer},
  {Scodeggio}, {Skillman}, {Solanes}, {Valotto}, {van Driel}, \& {van
  Zee}}]{Giovanelli2005}
{Giovanelli}, R., {et~al.} 2005, AJ, 130, 2598

\bibitem[{{Greisen}(2003)}]{Greisen2003}
{Greisen}, E.~W. 2003, Information Handling in Astronomy - Historical Vistas,
  285, 109

\bibitem[{{Haynes} {et~al.}(2011){Haynes}, {Giovanelli}, {Martin}, {Hess},
  {Saintonge}, {Adams}, {Hallenbeck}, {Hoffman}, {Huang}, {Kent}, {Koopmann},
  {Papastergis}, {Stierwalt}, {Balonek}, {Craig}, {Higdon}, {Kornreich},
  {Miller}, {O'Donoghue}, {Olowin}, {Rosenberg}, {Spekkens}, {Troischt}, \&
  {Wilcots}}]{Haynes2011}
{Haynes}, M.~P., {et~al.} 2011, AJ, 142, 170

\bibitem[{{Hibbard} {et~al.}(1994){Hibbard}, {Guhathakurta}, {van Gorkom}, \&
  {Schweizer}}]{Hibbard1994}
{Hibbard}, J.~E., {Guhathakurta}, P., {van Gorkom}, J.~H., \& {Schweizer}, F.
  1994, AJ, 107, 67

\bibitem[{{Hibbard} {et~al.}(2001){Hibbard}, {van der Hulst}, {Barnes}, \&
  {Rich}}]{Hibbard2001}
{Hibbard}, J.~E., {van der Hulst}, J.~M., {Barnes}, J.~E., \& {Rich}, R.~M.
  2001, AJ, 122, 2969

\bibitem[{Hopkins {et~al.}(2009)Hopkins, Somerville, Cox, Hernquist, Jogee,
  Kere\v{s}, Ma, Robertson, \& Stewart}]{Hopkins2009}
Hopkins, P.~F., {et~al.} 2009, MNRAS, 397, 802

\bibitem[{{Huang} {et~al.}(2012){Huang}, {Haynes}, {Giovanelli}, \&
  {Brinchmann}}]{Huang2012a}
{Huang}, S., {Haynes}, M.~P., {Giovanelli}, R., \& {Brinchmann}, J. 2012, ApJ,
  756, 113

\bibitem[{Ji {et~al.}(2014)Ji, Peirani, \& Yi}]{Ji2014}
Ji, I., Peirani, S., \& Yi, S.~K. 2014, A\&A

\bibitem[{{Jogee} {et~al.}(2009){Jogee}, {Miller}, {Penner}, {Skelton},
  {Conselice}, {Somerville}, {Bell}, {Zheng}, {Rix}, {Robaina}, {Barazza},
  {Barden}, {Borch}, {Beckwith}, {Caldwell}, {Peng}, {Heymans}, {McIntosh},
  {H{\"a}u{\ss}ler}, {Jahnke}, {Meisenheimer}, {Sanchez}, {Wisotzki}, {Wolf},
  \& {Papovich}}]{Jogee2009}
{Jogee}, S., {et~al.} 2009, ApJ, 697, 1971

\bibitem[{{Katz} {et~al.}(1996){Katz}, {Weinberg}, \& {Hernquist}}]{Katz1996}
{Katz}, N., {Weinberg}, D.~H., \& {Hernquist}, L. 1996, ApJS, 105, 19

\bibitem[{Kauffmann {et~al.}(2003)Kauffmann, Heckman, Tremonti, Brinchmann,
  Charlot, White, Ridgway, Brinkmann, Fukugita, Hall, Ivezi\'{c}, Richards, \&
  Schneider}]{Kauffmann2003}
Kauffmann, G., {et~al.} 2003, MNRAS, 346, 1055

\bibitem[{Kewley {et~al.}(2001)Kewley, Dopita, Sutherland, Heisler, \&
  Trevena}]{Kewley2001}
Kewley, L.~J., Dopita, M.~A., Sutherland, R.~S., Heisler, C.~A., \& Trevena, J.
  2001, ApJ, 556, 121

\bibitem[{{Kewley} {et~al.}(2006){Kewley}, {Groves}, {Kauffmann}, \&
  {Heckman}}]{Kewley2006}
{Kewley}, L.~J., {Groves}, B., {Kauffmann}, G., \& {Heckman}, T. 2006, MNRAS,
  372, 961

\bibitem[{{Khabiboulline} {et~al.}(2014){Khabiboulline}, {Steinhardt},
  {Silverman}, {Ellison}, {Mendel}, \& {Patton}}]{Khabiboulline2014}
{Khabiboulline}, E.~T., {Steinhardt}, C.~L., {Silverman}, J.~D., {Ellison},
  S.~L., {Mendel}, J.~T., \& {Patton}, D.~R. 2014, ApJ, 795, 62

\bibitem[{{Koss} {et~al.}(2011){Koss}, {Mushotzky}, {Veilleux}, {Winter},
  {Baumgartner}, {Tueller}, {Gehrels}, \& {Valencic}}]{Koss2011}
{Koss}, M., {Mushotzky}, R., {Veilleux}, S., {Winter}, L.~M., {Baumgartner},
  W., {Tueller}, J., {Gehrels}, N., \& {Valencic}, L. 2011, ApJ, 739, 57

\bibitem[{{Lambas} {et~al.}(2012){Lambas}, {Alonso}, {Mesa}, \&
  {O'Mill}}]{Lambas2012}
{Lambas}, D.~G., {Alonso}, S., {Mesa}, V., \& {O'Mill}, A.~L. 2012, A\&A, 539,
  A45

\bibitem[{{Lambas} {et~al.}(2003){Lambas}, {Tissera}, {Alonso}, \&
  {Coldwell}}]{Lambas2003}
{Lambas}, D.~G., {Tissera}, P.~B., {Alonso}, M.~S., \& {Coldwell}, G. 2003,
  MNRAS, 346, 1189

\bibitem[{{Lotz} {et~al.}(2010){Lotz}, {Jonsson}, {Cox}, \&
  {Primack}}]{Lotz2010}
{Lotz}, J.~M., {Jonsson}, P., {Cox}, T.~J., \& {Primack}, J.~R. 2010, MNRAS,
  404, 575

\bibitem[{{McMullin} {et~al.}(2007){McMullin}, {Waters}, {Schiebel}, {Young},
  \& {Golap}}]{McMullin2007}
{McMullin}, J.~P., {Waters}, B., {Schiebel}, D., {Young}, W., \& {Golap}, K.
  2007, in ASPCS, Vol. 376, Astronomical Data Analysis Software and Systems
  XVI, ed. R.~A. {Shaw}, F.~{Hill}, \& D.~J. {Bell}, 127

\bibitem[{{Mendel} {et~al.}(2014){Mendel}, {Simard}, {Palmer}, {Ellison}, \&
  {Patton}}]{Mendel2014}
{Mendel}, J.~T., {Simard}, L., {Palmer}, M., {Ellison}, S.~L., \& {Patton},
  D.~R. 2014, ApJS, 210, 3

\bibitem[{{Michel-Dansac} {et~al.}(2008){Michel-Dansac}, {Lambas}, {Alonso}, \&
  {Tissera}}]{Michel-Dansac2008}
{Michel-Dansac}, L., {Lambas}, D.~G., {Alonso}, M.~S., \& {Tissera}, P. 2008,
  MNRAS, 386, L82

\bibitem[{{Mihos} \& {Hernquist}(1994)}]{Mihos1994}
{Mihos}, J.~C., \& {Hernquist}, L. 1994, ApJL, 425, L13

\bibitem[{{Mihos} \& {Hernquist}(1996)}]{Mihos1996}
---. 1996, ApJ, 464, 641

\bibitem[{{Mihos} {et~al.}(1992){Mihos}, {Richstone}, \& {Bothun}}]{Mihos1992}
{Mihos}, J.~C., {Richstone}, D.~O., \& {Bothun}, G.~D. 1992, ApJ, 400, 153

\bibitem[{{Montuori} {et~al.}(2010){Montuori}, {Di Matteo}, {Lehnert},
  {Combes}, \& {Semelin}}]{Montuori2010}
{Montuori}, M., {Di Matteo}, P., {Lehnert}, M.~D., {Combes}, F., \& {Semelin},
  B. 2010, A\&A, 518, A56

\bibitem[{{Moreno} {et~al.}(2014){Moreno}, {Torrey}, {Ellison}, {Patton},
  {Bluck}, {Bansal}, \& {Hernquist}}]{Moreno2014}
{Moreno}, J., {Torrey}, P., {Ellison}, S.~L., {Patton}, D.~R., {Bluck},
  A.~F.~L., {Bansal}, G., \& {Hernquist}, L. 2014, MNRAS, submitted

\bibitem[{{Patton} {et~al.}(2011){Patton}, {Ellison}, {Simard}, {McConnachie},
  \& {Mendel}}]{Patton2011}
{Patton}, D.~R., {Ellison}, S.~L., {Simard}, L., {McConnachie}, A.~W., \&
  {Mendel}, J.~T. 2011, MNRAS, 412, 591

\bibitem[{{Patton} {et~al.}(2013){Patton}, {Torrey}, {Ellison}, {Mendel}, \&
  {Scudder}}]{Patton2013}
{Patton}, D.~R., {Torrey}, P., {Ellison}, S.~L., {Mendel}, J.~T., \& {Scudder},
  J.~M. 2013, MNRAS, 433, L59

\bibitem[{{Perez} {et~al.}(2011){Perez}, {Michel-Dansac}, \&
  {Tissera}}]{Perez2011a}
{Perez}, J., {Michel-Dansac}, L., \& {Tissera}, P.~B. 2011, MNRAS, 417, 580

\bibitem[{{Perez} {et~al.}(2009){Perez}, {Tissera}, \& {Blaizot}}]{Perez2009}
{Perez}, J., {Tissera}, P., \& {Blaizot}, J. 2009, MNRAS, 397, 748

\bibitem[{{Perret} {et~al.}(2014){Perret}, {Renaud}, {Epinat}, {Amram},
  {Bournaud}, {Contini}, {Teyssier}, \& {Lambert}}]{Perret2014}
{Perret}, V., {Renaud}, F., {Epinat}, B., {Amram}, P., {Bournaud}, F.,
  {Contini}, T., {Teyssier}, R., \& {Lambert}, J.-C. 2014, A\&A, 562, A1

\bibitem[{{Robaina} {et~al.}(2009){Robaina}, {Bell}, {Skelton}, {McIntosh},
  {Somerville}, {Zheng}, {Rix}, {Bacon}, {Balogh}, {Barazza}, {Barden},
  {B{\"o}hm}, {Caldwell}, {Gallazzi}, {Gray}, {H{\"a}ussler}, {Heymans},
  {Jahnke}, {Jogee}, {van Kampen}, {Lane}, {Meisenheimer}, {Papovich}, {Peng},
  {S{\'a}nchez}, {Skibba}, {Taylor}, {Wisotzki}, \& {Wolf}}]{Robaina2009}
{Robaina}, A.~R., {et~al.} 2009, ApJ, 704, 324

\bibitem[{{Sabater} {et~al.}(2013){Sabater}, {Best}, \&
  {Argudo-Fern{\'a}ndez}}]{Sabater2013}
{Sabater}, J., {Best}, P.~N., \& {Argudo-Fern{\'a}ndez}, M. 2013, MNRAS, 430,
  638

\bibitem[{{Saintonge} {et~al.}(2011){Saintonge}, {Kauffmann}, {Kramer},
  {Tacconi}, {Buchbender}, {Catinella}, {Fabello}, {Graci{\'a}-Carpio}, {Wang},
  {Cortese}, {Fu}, {Genzel}, {Giovanelli}, {Guo}, {Haynes}, {Heckman},
  {Krumholz}, {Lemonias}, {Li}, {Moran}, {Rodriguez-Fernandez}, {Schiminovich},
  {Schuster}, \& {Sievers}}]{Saintonge2011}
{Saintonge}, A., {et~al.} 2011, MNRAS, 415, 32

\bibitem[{{Saintonge} {et~al.}(2012){Saintonge}, {Tacconi}, {Fabello}, {Wang},
  {Catinella}, {Genzel}, {Graci{\'a}-Carpio}, {Kramer}, {Moran}, {Heckman},
  {Schiminovich}, {Schuster}, \& {Wuyts}}]{Saintonge2012}
---. 2012, ApJ, 758, 73

\bibitem[{{Salim} {et~al.}(2007){Salim}, {Rich}, {Charlot}, {Brinchmann},
  {Johnson}, {Schiminovich}, {Seibert}, {Mallery}, {Heckman}, {Forster},
  {Friedman}, {Martin}, {Morrissey}, {Neff}, {Small}, {Wyder}, {Bianchi},
  {Donas}, {Lee}, {Madore}, {Milliard}, {Szalay}, {Welsh}, \& {Yi}}]{Salim2007}
{Salim}, S., {et~al.} 2007, ApJS, 173, 267

\bibitem[{{Satyapal} {et~al.}(2014){Satyapal}, {Ellison}, {McAlpine}, {Hickox},
  {Patton}, \& {Mendel}}]{Satyapal2014b}
{Satyapal}, S., {Ellison}, S.~L., {McAlpine}, W., {Hickox}, R.~C., {Patton},
  D.~R., \& {Mendel}, J.~T. 2014, MNRAS, 441, 1297

\bibitem[{{Scudder} {et~al.}(2012{\natexlab{a}}){Scudder}, {Ellison}, \&
  {Mendel}}]{Scudder2012}
{Scudder}, J.~M., {Ellison}, S.~L., \& {Mendel}, J.~T. 2012{\natexlab{a}},
  MNRAS, 423, 2690

\bibitem[{{Scudder} {et~al.}(2012{\natexlab{b}}){Scudder}, {Ellison}, {Torrey},
  {Patton}, \& {Mendel}}]{Scudder2012b}
{Scudder}, J.~M., {Ellison}, S.~L., {Torrey}, P., {Patton}, D.~R., \& {Mendel},
  J.~T. 2012{\natexlab{b}}, MNRAS, 426, 549

\bibitem[{{Sengupta} {et~al.}(2013){Sengupta}, {Dwarakanath}, {Saikia}, \&
  {Scott}}]{Sengupta2013}
{Sengupta}, C., {Dwarakanath}, K.~S., {Saikia}, D.~J., \& {Scott}, T.~C. 2013,
  MNRAS, 431, L1

\bibitem[{{Silverman} {et~al.}(2011){Silverman}, {Kampczyk}, {Jahnke},
  {Andrae}, {Lilly}, {Elvis}, {Civano}, {Mainieri}, {Vignali}, {Zamorani},
  {Nair}, {Le F{\`e}vre}, {de Ravel}, {Bardelli}, {Bongiorno}, {Bolzonella},
  {Cappi}, {Caputi}, {Carollo}, {Contini}, {Coppa}, {Cucciati}, {de la Torre},
  {Franzetti}, {Garilli}, {Halliday}, {Hasinger}, {Iovino}, {Knobel},
  {Koekemoer}, {Kova{\v c}}, {Lamareille}, {Le Borgne}, {Le Brun}, {Maier},
  {Mignoli}, {Pello}, {P{\'e}rez-Montero}, {Ricciardelli}, {Peng}, {Scodeggio},
  {Tanaka}, {Tasca}, {Tresse}, {Vergani}, {Zucca}, {Brusa}, {Cappelluti},
  {Comastri}, {Finoguenov}, {Fu}, {Gilli}, {Hao}, {Ho}, \&
  {Salvato}}]{Silverman2011}
{Silverman}, J.~D., {et~al.} 2011, ApJ, 743, 2

\bibitem[{{Simard} {et~al.}(2011){Simard}, {Mendel}, {Patton}, {Ellison}, \&
  {McConnachie}}]{Simard2011}
{Simard}, L., {Mendel}, J.~T., {Patton}, D.~R., {Ellison}, S.~L., \&
  {McConnachie}, A.~W. 2011, ApJS, 196, 11

\bibitem[{{Smith}(1991)}]{Smith1991}
{Smith}, B.~J. 1991, ApJ, 378, 39

\bibitem[{{Smith}(1994)}]{Smith1994a}
---. 1994, AJ, 107, 1695

\bibitem[{{Springel}(2005)}]{Springel2005a}
{Springel}, V. 2005, MNRAS, 364, 1105

\bibitem[{{Springel} \& {Hernquist}(2003)}]{Springel2003}
{Springel}, V., \& {Hernquist}, L. 2003, MNRAS, 339, 289

\bibitem[{{Stark} {et~al.}(2013){Stark}, {Kannappan}, {Wei}, {Baker}, {Leroy},
  {Eckert}, \& {Vogel}}]{Stark2013}
{Stark}, D.~V., {Kannappan}, S.~J., {Wei}, L.~H., {Baker}, A.~J., {Leroy},
  A.~K., {Eckert}, K.~D., \& {Vogel}, S.~N. 2013, ApJ, 769, 82

\bibitem[{Stierwalt {et~al.}(2014)Stierwalt, Besla, Patton, Johnson,
  Kallivayalil, Putman, Privon, \& Ross}]{Stierwalt2014}
Stierwalt, S., Besla, G., Patton, D., Johnson, K., Kallivayalil, N., Putman,
  M., Privon, G., \& Ross, G. 2014

\bibitem[{{Torrey} {et~al.}(2012){Torrey}, {Cox}, {Kewley}, \&
  {Hernquist}}]{Torrey2012}
{Torrey}, P., {Cox}, T.~J., {Kewley}, L., \& {Hernquist}, L. 2012, ApJ, 746,
  108

\bibitem[{Whitaker {et~al.}(2014)Whitaker, Rigby, Brammer, Gladders, Sharon,
  Teng, \& Wuyts}]{Whitaker2014}
Whitaker, K.~E., Rigby, J.~R., Brammer, G.~B., Gladders, M.~D., Sharon, K.,
  Teng, S.~H., \& Wuyts, E. 2014, ApJ

\bibitem[{{Wong} {et~al.}(2011){Wong}, {Blanton}, {Burles}, {Coil}, {Cool},
  {Eisenstein}, {Moustakas}, {Zhu}, \& {Arnouts}}]{Wong2011}
{Wong}, K.~C., {et~al.} 2011, ApJ, 728, 119

\bibitem[{{Woods} \& {Geller}(2007)}]{Woods2007}
{Woods}, D.~F., \& {Geller}, M.~J. 2007, AJ, 134, 527

\bibitem[{{Xu} {et~al.}(2010){Xu}, {Domingue}, {Cheng}, {Lu}, {Huang}, {Gao},
  {Mazzarella}, {Cutri}, {Sun}, \& {Surace}}]{Xu2010}
{Xu}, C.~K., {et~al.} 2010, ApJ, 713, 330

\bibitem[{{Xu} {et~al.}(2012){Xu}, {Shupe}, {B{\'e}thermin}, {Aussel}, {Berta},
  {Bock}, {Bridge}, {Conley}, {Cooray}, {Elbaz}, {Franceschini}, {Le Floc'h},
  {Lu}, {Lutz}, {Magnelli}, {Marsden}, {Oliver}, {Pozzi}, {Riguccini},
  {Schulz}, {Scoville}, {Vaccari}, {Vieira}, {Wang}, \& {Zemcov}}]{Xu2012}
---. 2012, ApJ, 760, 72

\bibitem[{{Zasov} \& {Sulentic}(1994)}]{Zasov1994}
{Zasov}, A.~V., \& {Sulentic}, J.~W. 1994, ApJ, 430, 179

\end{thebibliography}
\appendix
\section{Additional Figures}
\begin{figure}
   \centering
   \includegraphics[width=250px]{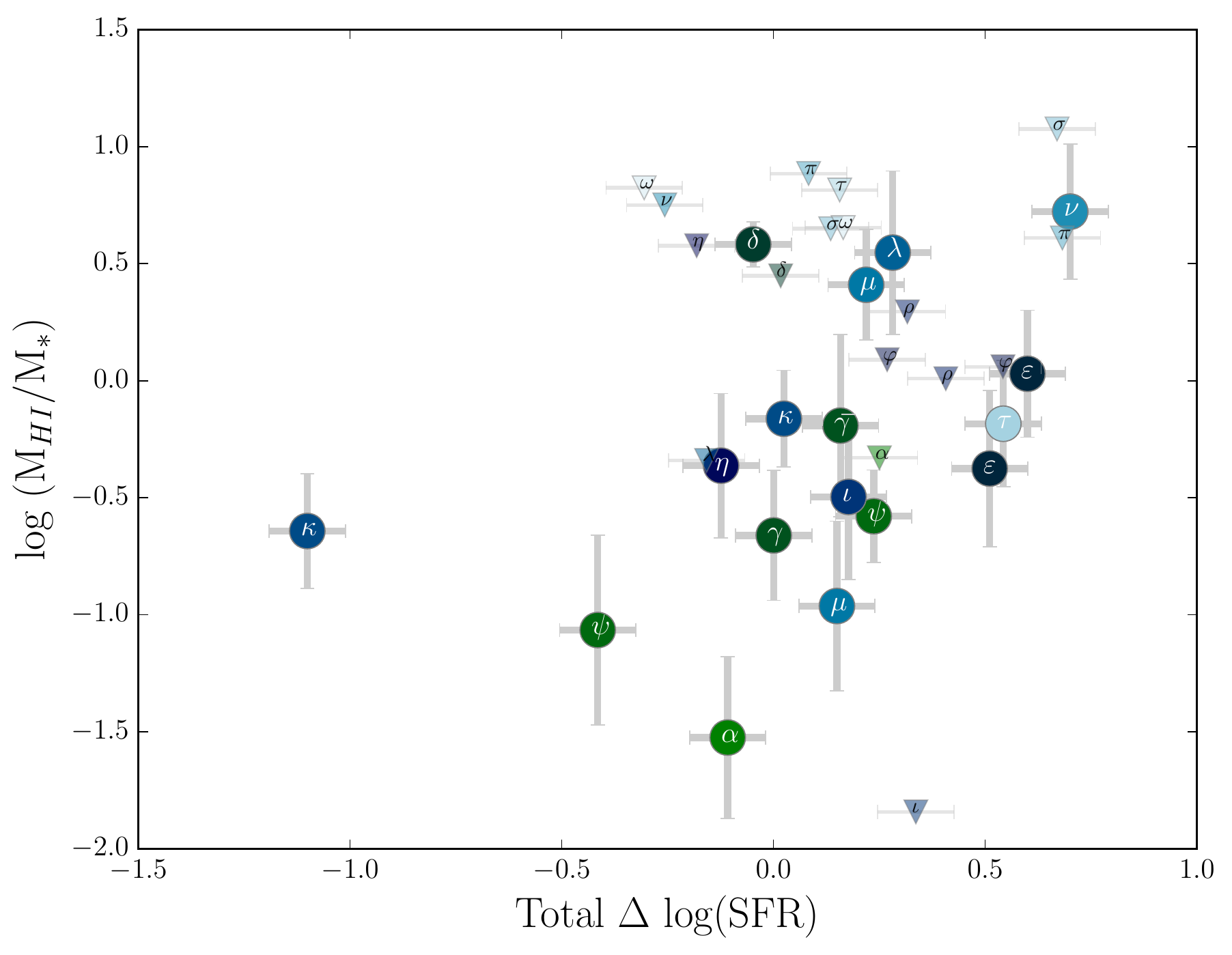}
   \caption[~Gas fraction \& \delsfr]{The log of the gas fraction vs. \delsfr.  Points are colour-coded and labeled with a greek letter according to pair.  Large circles indicate S/N $>3.0$, with smaller triangles indicating the $3\sigma$ upper limits for the maximum amount of flux that could exist in the spectrum.}
   \label{fig:UL_gf_tsfr}
\end{figure}

\begin{figure*}
   \centering
   \begin{minipage}[c]{8.5cm}
   \includegraphics[height=245px]{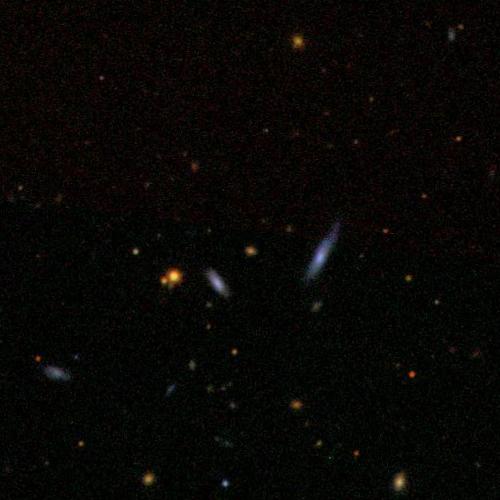} 
   \end{minipage}
   \begin{minipage}[c]{8.5cm}
   \includegraphics[height=250px]{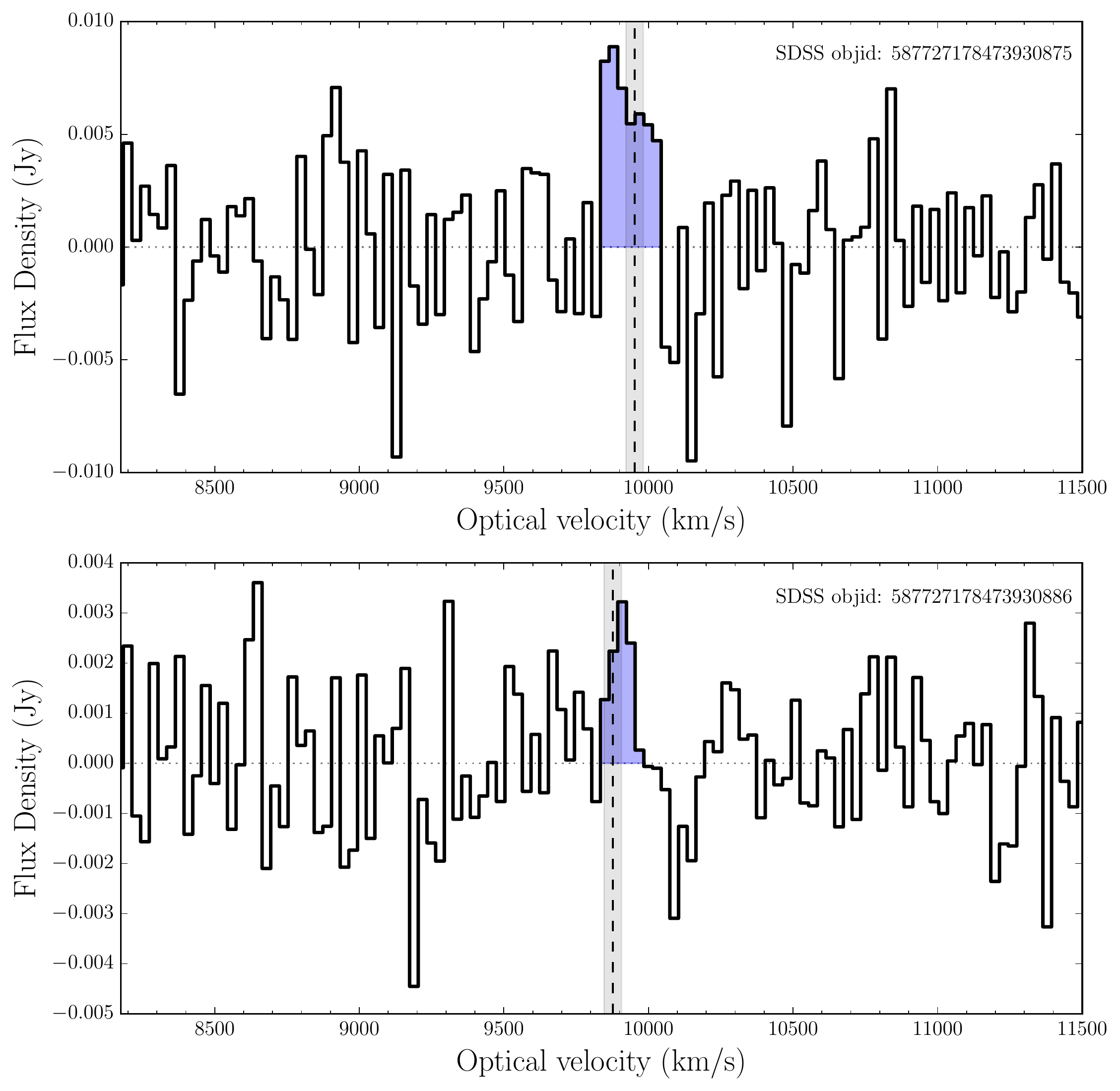}
    \end{minipage}
    \caption[~SDSS: 588018056204780081 \& 588018056204780049]{Left panel: SDSS thumbnail of galaxy pair 587727178473930875 (right) \& 587727178473930886 (left). Right panel: Flux density vs. optical velocity for galaxy pair 587727178473930875 (upper) \& 587727178473930886 (lower).  The vertical dashed line indicates the velocity of the galaxy, horizontal dotted line indicates 0 flux.  The blue shaded region indicates the region of the spectra used for signal to noise and gas mass calculations. The grey shaded region around the indicated velocity of the galaxy indicates the uncertainty of the SDSS redshift ($\pm 30$ \kms). The upper spectrum has a peak/RMS S/N of 2.62, while the lower spectrum has a peak/RMS S/N of 2.24.\label{fig:0875}}
\end{figure*}

\begin{figure*}
   \centering
   \begin{minipage}[c]{8.5cm}
   \includegraphics[height=245px]{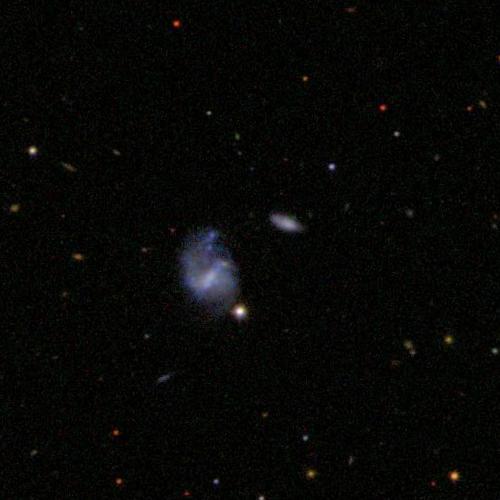} 
   \end{minipage}
   \begin{minipage}[c]{8.5cm}
   \includegraphics[height=250px]{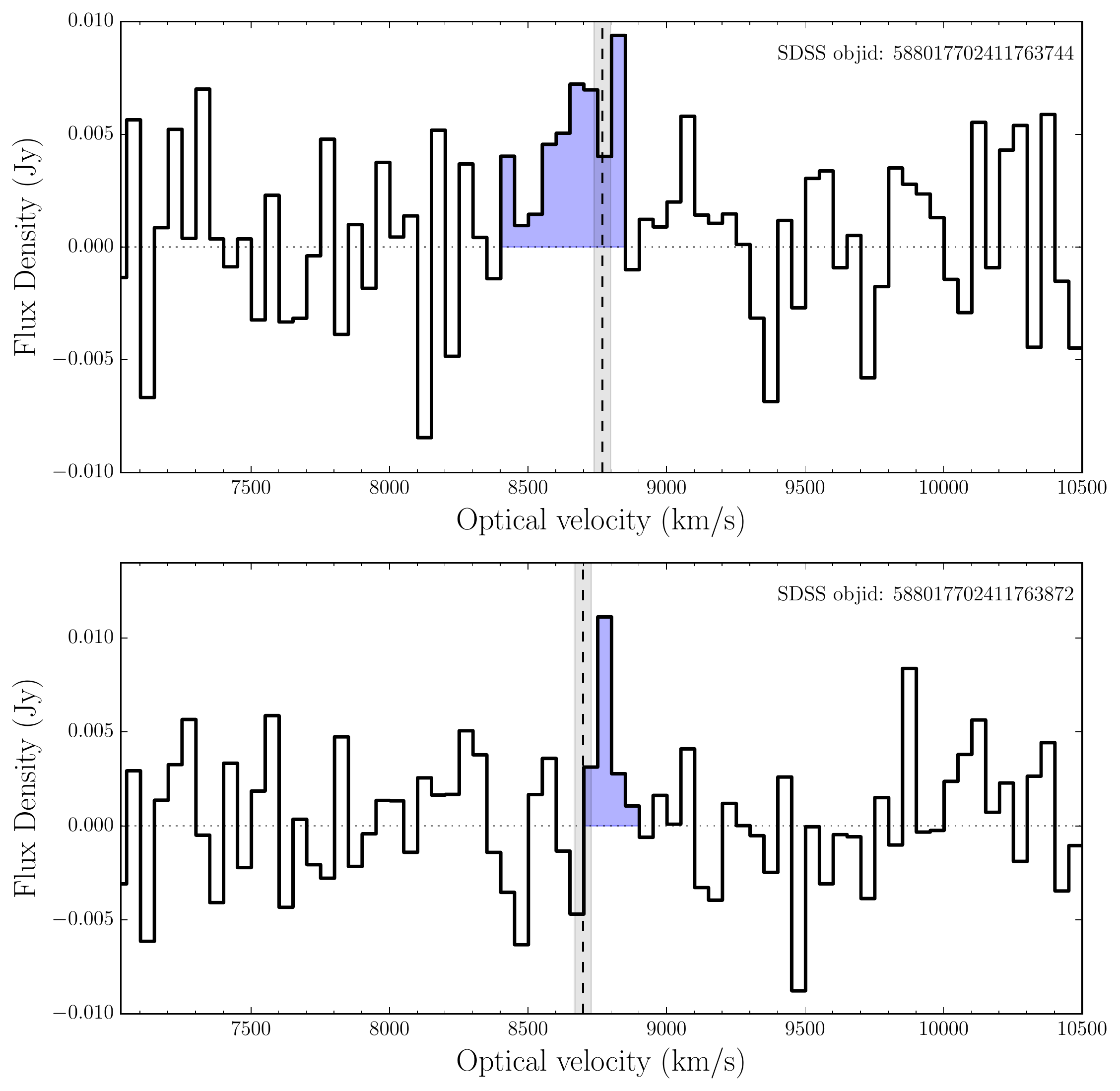}
    \end{minipage}
    \caption{Same as in Figure \ref{fig:0875}.  Left panel: SDSS thumbnail of galaxy pair 588017702411763744 (lower left) \& 588017702411763872 (upper right). 
    Right panel: Flux density vs. optical velocity for galaxy pair 588017702411763744 (upper) \& 588017702411763872 (lower). The upper spectrum has a peak/RMS S/N of 2.67, while the lower spectrum has a peak/RMS S/N of 2.79. \label{fig:3744}}
\end{figure*}

\begin{figure*}
   \centering
   \begin{minipage}[c]{8.5cm}
   \includegraphics[height=245px]{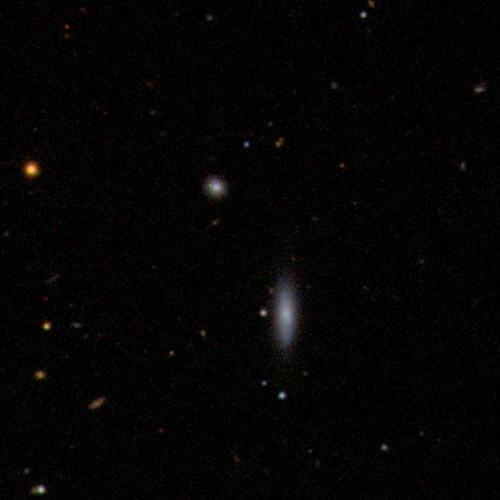} 
   \end{minipage}
   \begin{minipage}[c]{8.5cm}
   \includegraphics[height=250px]{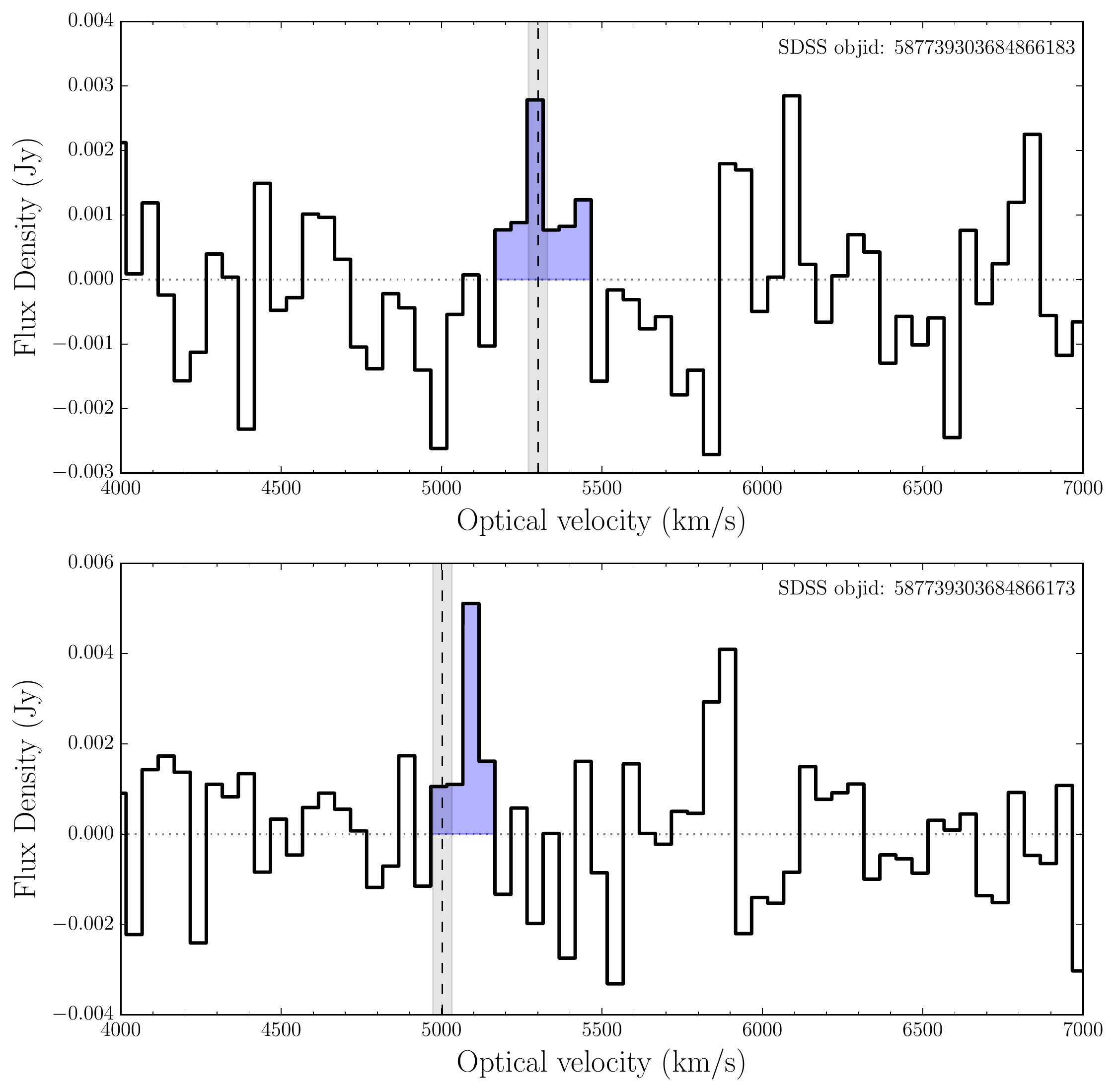}
    \end{minipage}
    \caption{Same as in Figure \ref{fig:0875}.  Left panel: SDSS thumbnail of galaxy pair 587739303684866183 (upper left) \& 587739303684866173 (lower right).
    Right panel: Flux density vs. optical velocity for galaxy pair 587739303684866183 (upper) \& 587739303684866173 (lower). The upper spectrum has a peak/RMS S/N of 2.35, while the lower spectrum has a peak/RMS S/N of 3.57. 
    \label{fig:04398}}
\end{figure*}

\begin{figure*}
   \centering
   \begin{minipage}[c]{8.5cm}
   \includegraphics[height=245px]{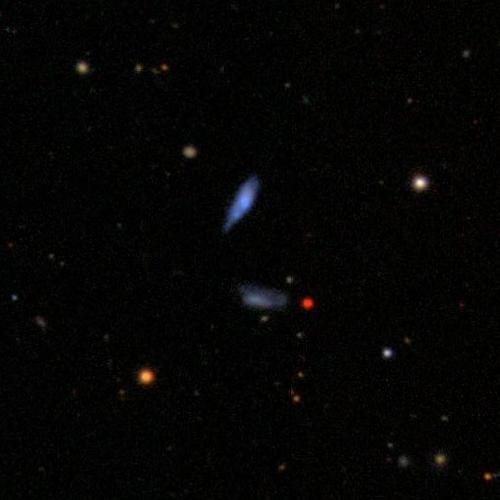} 
   \end{minipage}
   \begin{minipage}[c]{8.5cm}
   \includegraphics[height=250px]{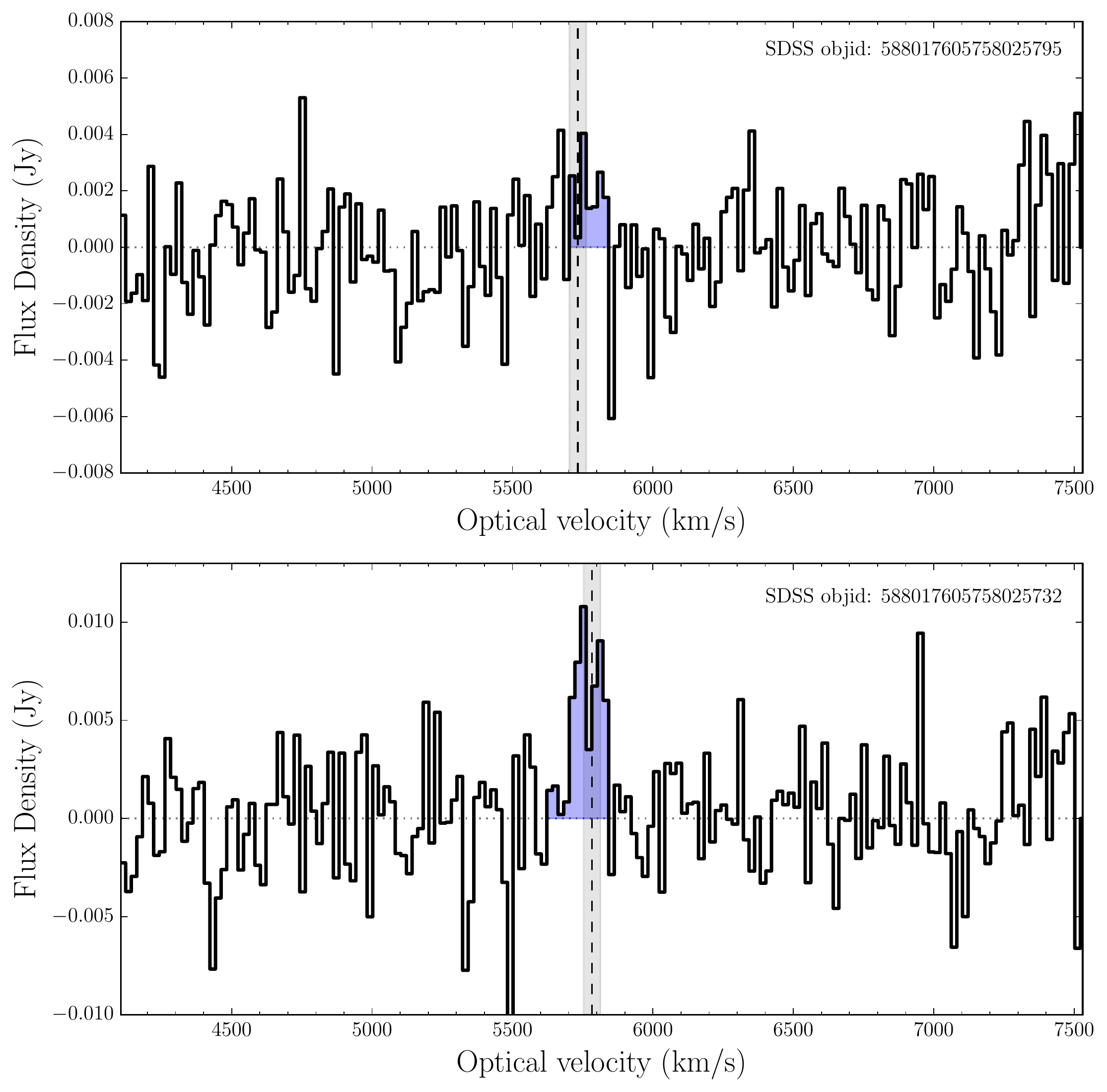}
    \end{minipage}
    \caption{Same as in Figure \ref{fig:0875}.  Left panel: SDSS thumbnail of galaxy pair 588017605758025795 (lower) \& 588017605758025732 (upper).
    Right panel: Flux density vs. optical velocity for galaxy pair 588017605758025795 (upper) \& 588017605758025732 (lower). The upper spectrum has a peak/RMS S/N of 1.93, while the lower spectrum has a peak/RMS S/N of 3.62. 
    \label{fig:5795}}
\end{figure*}

\begin{figure*}
   \centering
   \begin{minipage}[c]{8.5cm}
   \includegraphics[height=245px]{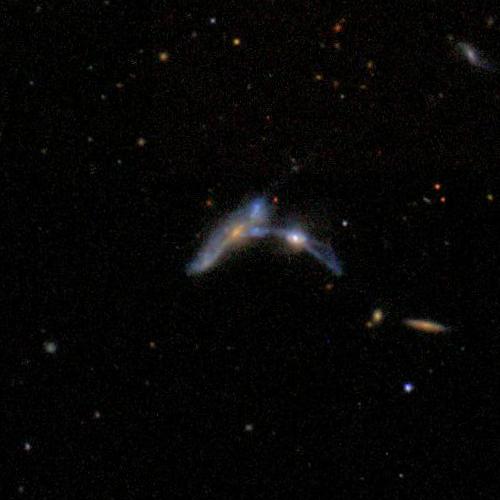} 
   \end{minipage}
   \begin{minipage}[c]{8.5cm}
   \includegraphics[height=250px]{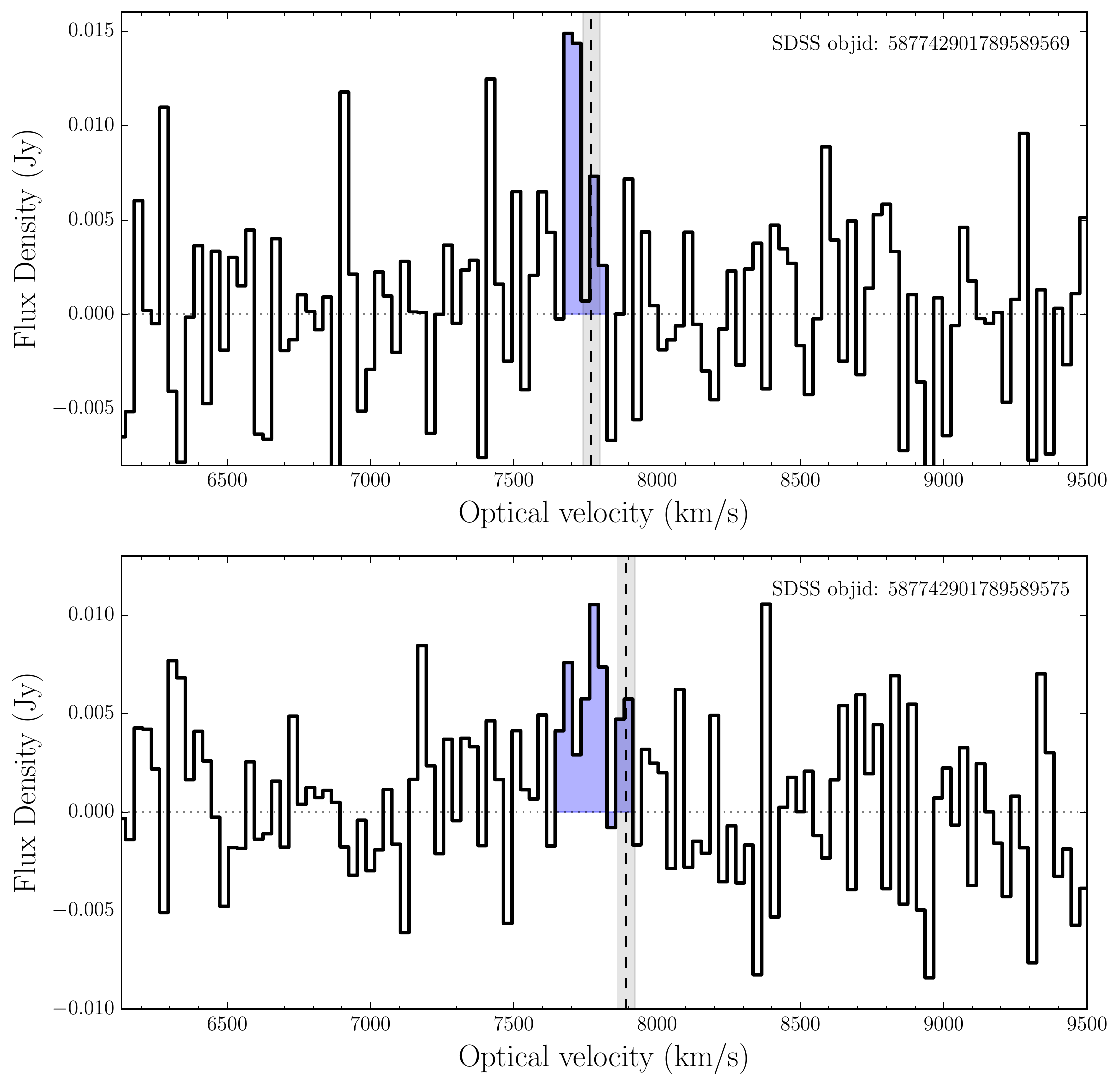}
    \end{minipage}
    \caption{Same as in Figure \ref{fig:0875}.  Left panel: SDSS thumbnail of galaxy pair 587742901789589569 (right) \& 587742901789589575 (left).
    Right panel: Flux density vs. optical velocity for galaxy pair 587742901789589569 (upper) \& 587742901789589575 (lower). The upper spectrum has a peak/RMS S/N of 3.20, while the lower spectrum has a peak/RMS S/N of 2.70.
    \label{fig:9569}}
\end{figure*}

\begin{figure*}
   \centering
   \begin{minipage}[c]{8.5cm}
   \includegraphics[height=245px]{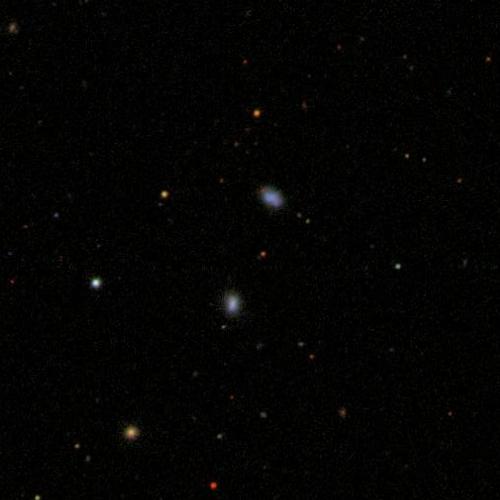} 
   \end{minipage}
   \begin{minipage}[c]{8.5cm}
   \includegraphics[height=250px]{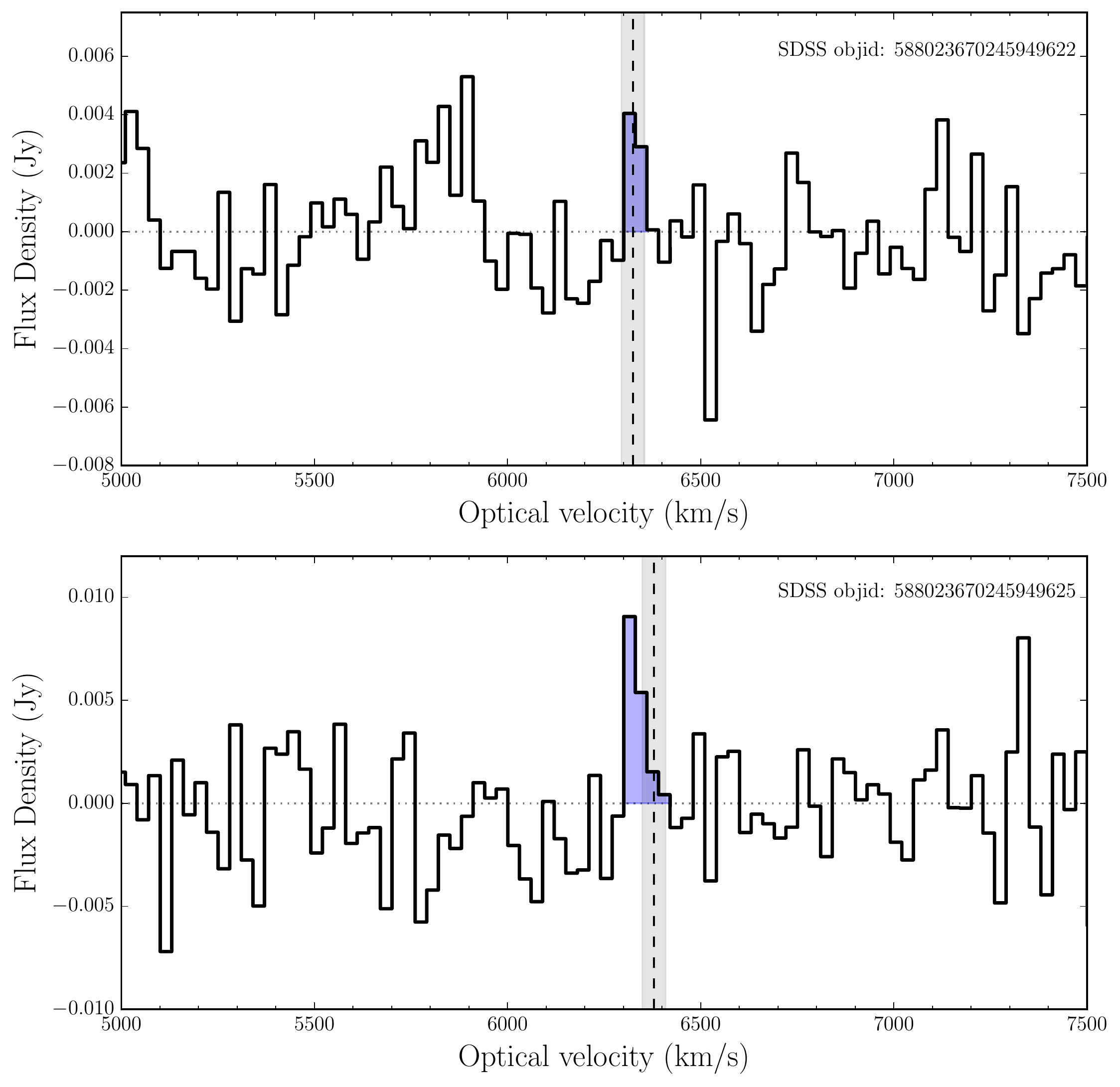}
    \end{minipage}
    \caption{Same as in Figure \ref{fig:0875}.  Left panel: SDSS thumbnail of galaxy pair 588023670245949622 (upper) \& 588023670245949625 (lower).
    Right panel: Flux density vs. optical velocity for galaxy pair 588023670245949622 (upper) \& 588023670245949625 (lower). The upper spectrum has a peak/RMS S/N of 1.61, while the lower spectrum has a peak/RMS S/N of 2.73.
    \label{fig:9622}}
\end{figure*}

\begin{figure*}
   \centering
   \begin{minipage}[c]{8.5cm}
   \includegraphics[height=245px]{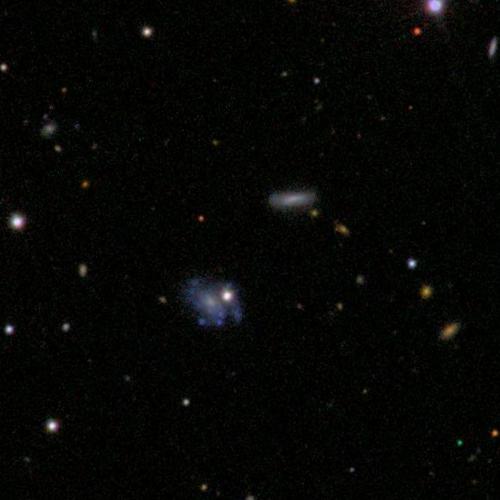} 
   \end{minipage}
   \begin{minipage}[c]{8.5cm}
   \includegraphics[height=250px]{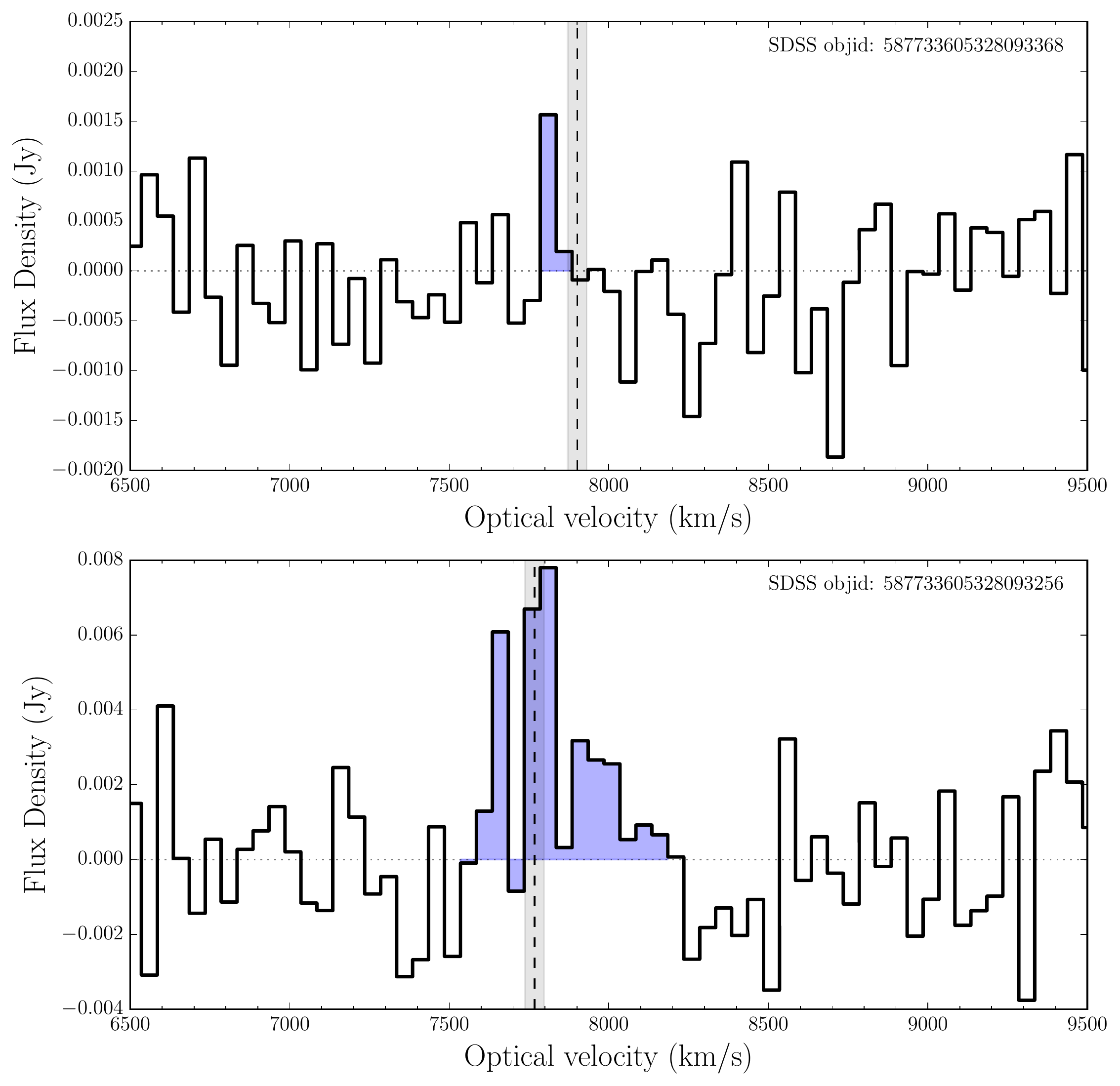}
    \end{minipage}
    \caption{Same as in Figure \ref{fig:0875}.  Left panel: SDSS thumbnail of galaxy pair 587733605328093368 (upper right) \& 587733605328093256 (lower left).
    Right panel: Flux density vs. optical velocity for galaxy pair 587733605328093368 (upper) \& 587733605328093256 (lower). The upper spectrum has a peak/RMS S/N of 2.26, while the lower spectrum has a peak/RMS S/N of 4.04. 
    \label{fig:21528}}
\end{figure*}

\begin{figure*}
   \centering
   \begin{minipage}[c]{8.5cm}
   \includegraphics[height=245px]{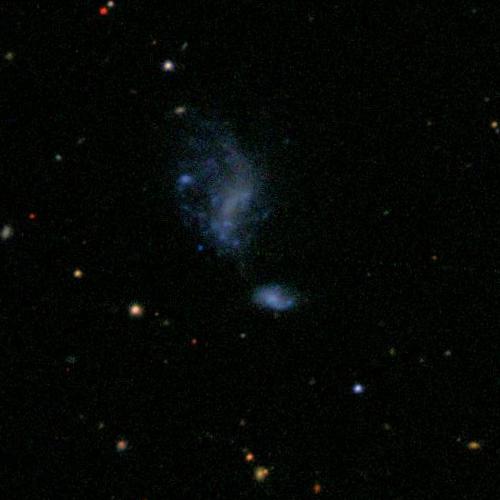} 
   \end{minipage}
   \begin{minipage}[c]{8.5cm}
   \includegraphics[height=250px]{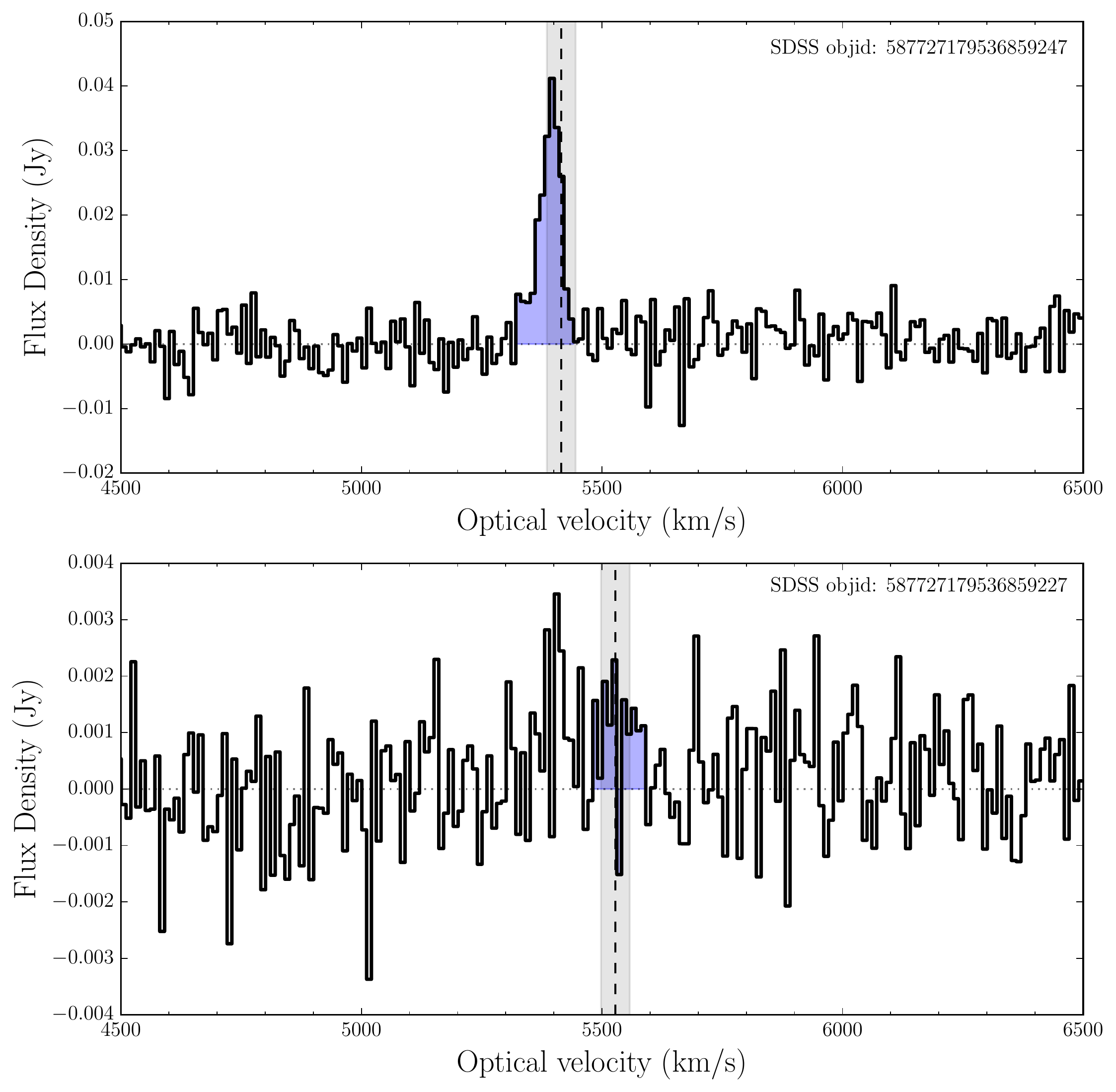}
    \end{minipage}
    \caption{Same as in Figure \ref{fig:0875}.  Left panel: SDSS thumbnail of galaxy pair 587727179536859247 (upper left) \& 587727179536859227 (lower right).
    Right panel: Flux density vs. optical velocity for galaxy pair 587727179536859247 (upper) \& 587727179536859227 (lower). The upper spectrum has a peak/RMS S/N of 11.40, while the lower spectrum has a peak/RMS S/N of 2.02.
    \label{fig:25232}}
\end{figure*}

\begin{figure*}
   \centering
   \begin{minipage}[c]{8.5cm}
   \includegraphics[height=245px]{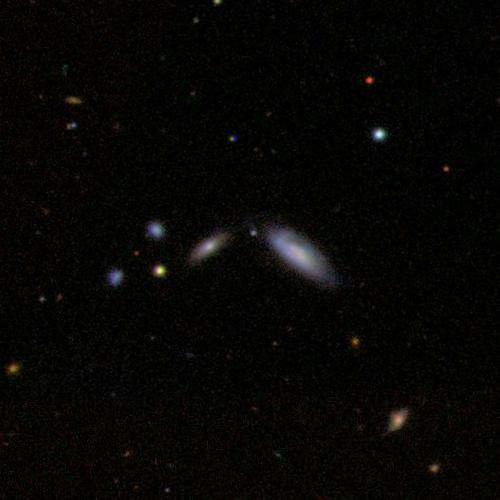} 
   \end{minipage}
   \begin{minipage}[c]{8.5cm}
   \includegraphics[height=250px]{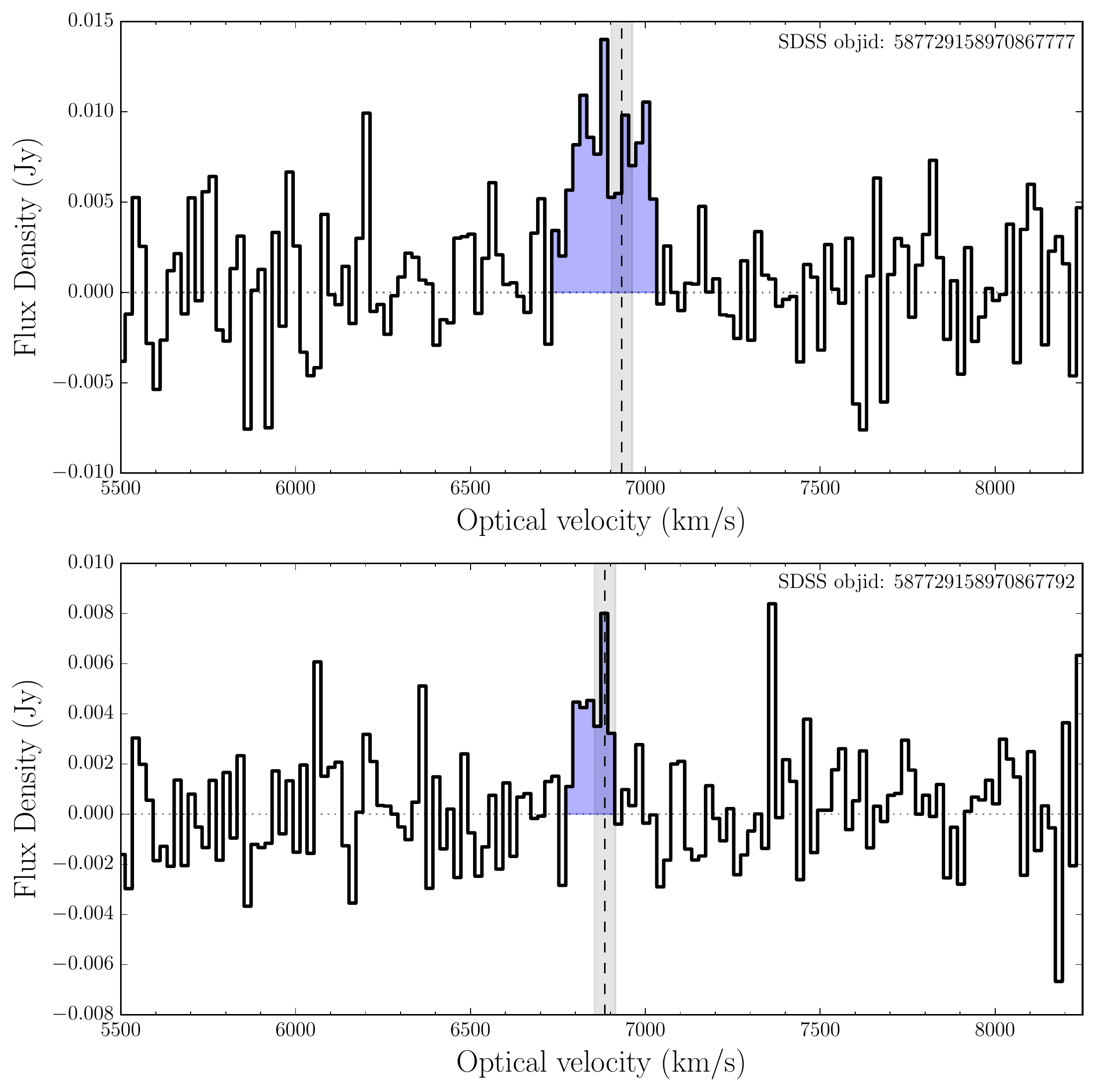}
    \end{minipage}
    \caption{Same as in Figure \ref{fig:0875}.  Left panel: SDSS thumbnail of galaxy pair 587729158970867777 (right) \& 587729158970867792 (left).
    Right panel: Flux density vs. optical velocity for galaxy pair 587729158970867777 (upper) \& 587729158970867792 (lower). The upper spectrum has a peak/RMS S/N of 4.09, while the lower spectrum has a peak/RMS S/N of 3.71.
    \label{fig:31481}}
\end{figure*}

\begin{figure*}
   \centering
   \begin{minipage}[c]{8.5cm}
   \includegraphics[height=245px]{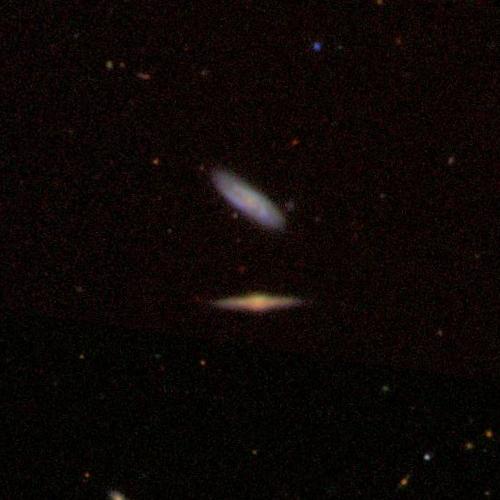} 
   \end{minipage}
   \begin{minipage}[c]{8.5cm}
   \includegraphics[height=250px]{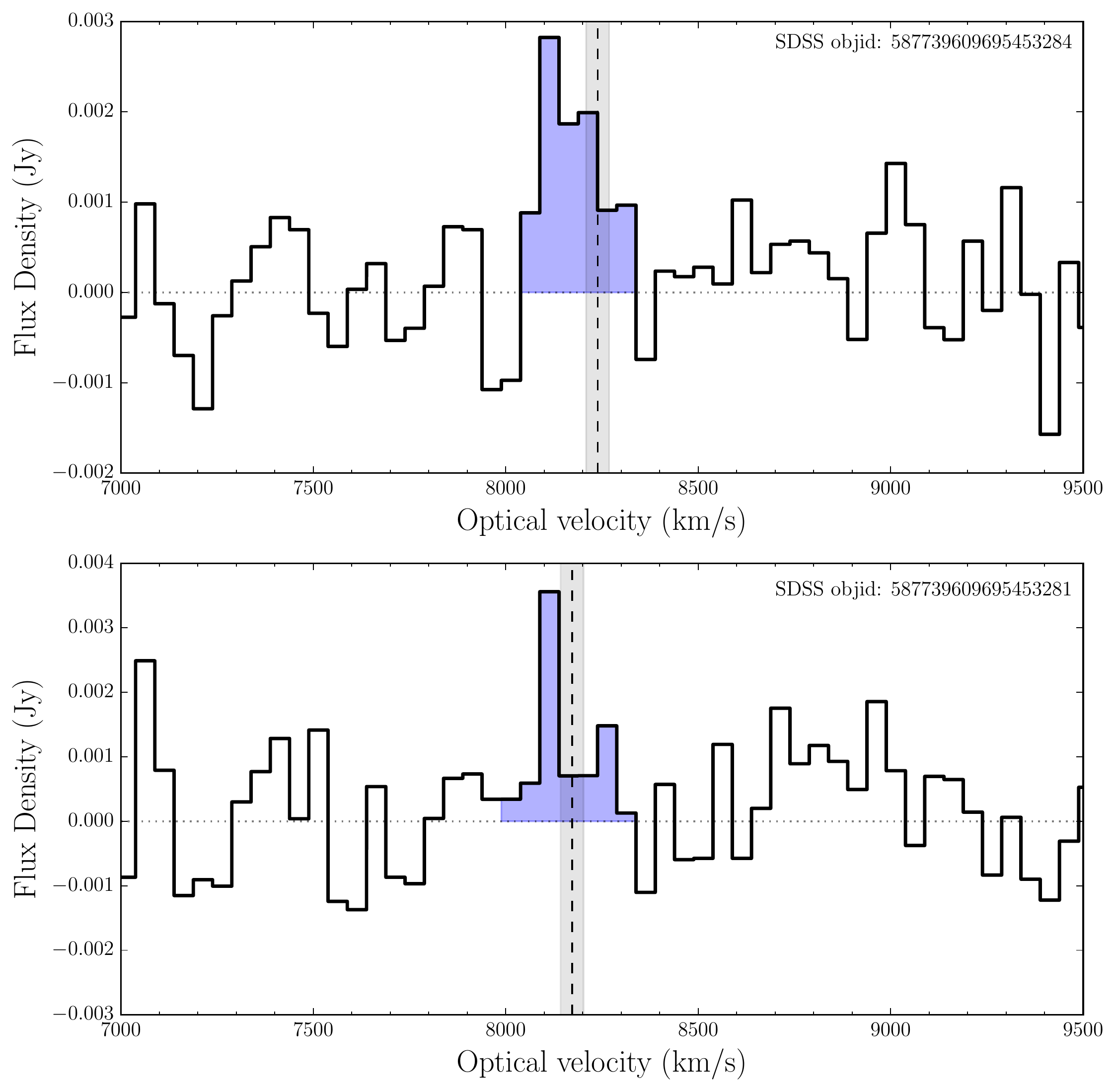}
    \end{minipage}
    \caption{Same as in Figure \ref{fig:0875}.  Left panel: SDSS thumbnail of galaxy pair 587739609695453284 (upper) \& 587739609695453281 (lower).
    Right panel: Flux density vs. optical velocity for galaxy pair 587739609695453284 (upper) \& 587739609695453281 (lower). The upper spectrum has a peak/RMS S/N of 4.08, while the lower spectrum has a peak/RMS S/N of 3.69.
    \label{fig:37037}}
\end{figure*}

\begin{figure*}
   \centering
   \begin{minipage}[c]{8.5cm}
   \includegraphics[height=245px]{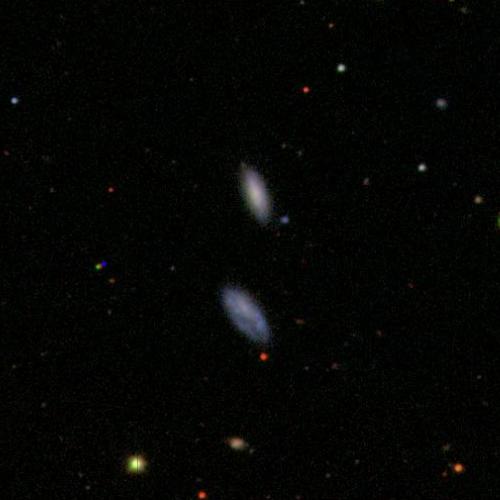} 
   \end{minipage}
   \begin{minipage}[c]{8.5cm}
   \includegraphics[height=250px]{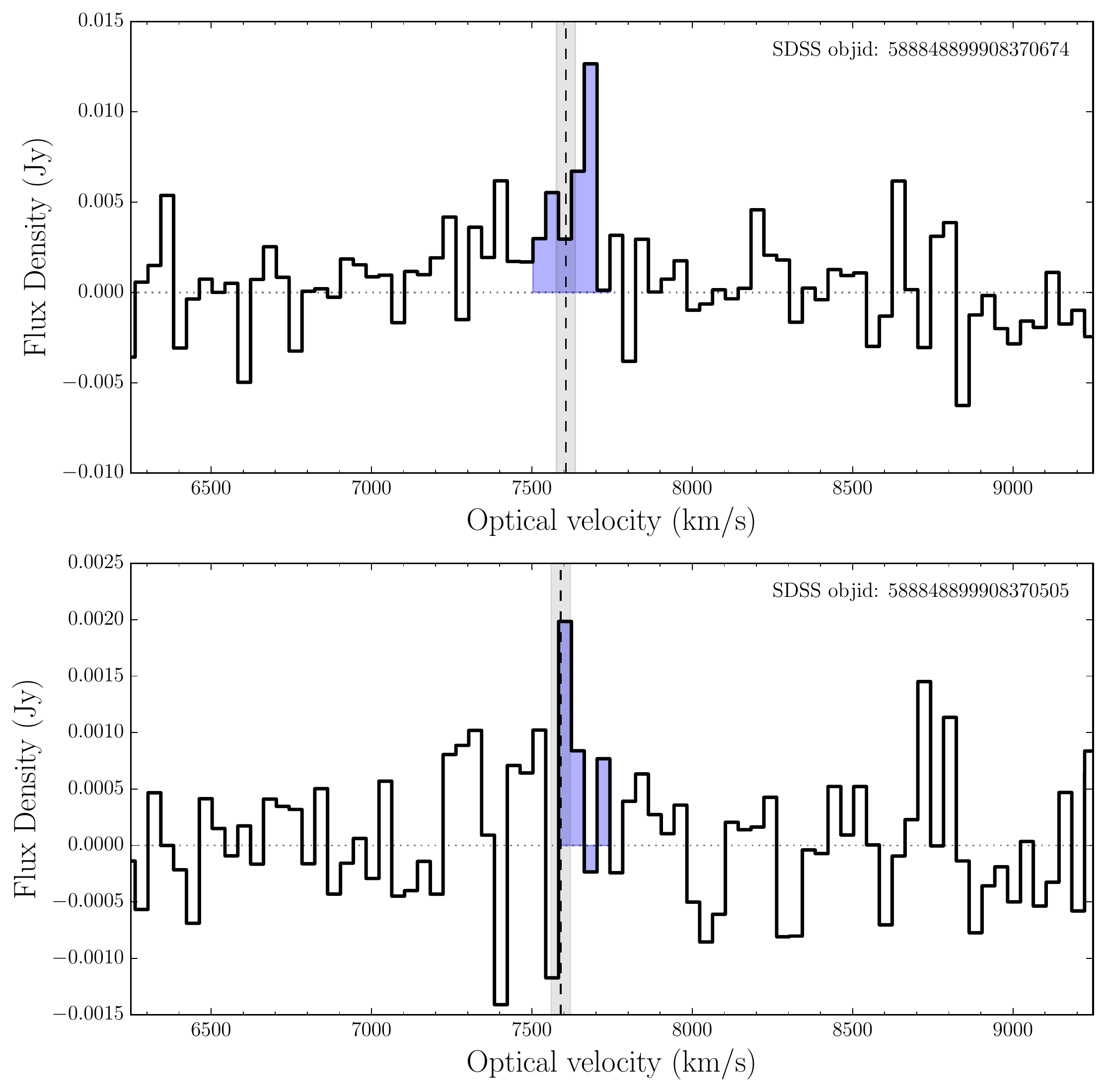}
    \end{minipage}
    \caption{Same as in Figure \ref{fig:0875}.  Left panel: SDSS thumbnail of galaxy pair 588848899908370674 (lower) \& 588848899908370505 (upper).
    Right panel: Flux density vs. optical velocity for galaxy pair 588848899908370674 (upper) \& 588848899908370505 (lower).  The upper spectrum has a peak/RMS S/N of 4.54, while the lower spectrum has a peak/RMS S/N of 3.26.
    \label{fig:62037}}
\end{figure*}

\begin{figure*}
   \centering
   \begin{minipage}[c]{8.5cm}
   \includegraphics[height=245px]{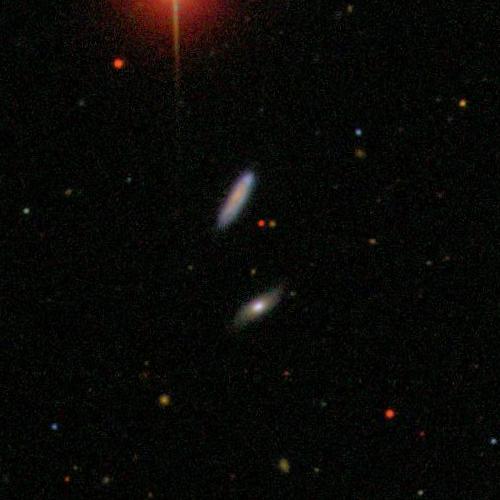} 
   \end{minipage}
   \begin{minipage}[c]{8.5cm}
   \includegraphics[height=250px]{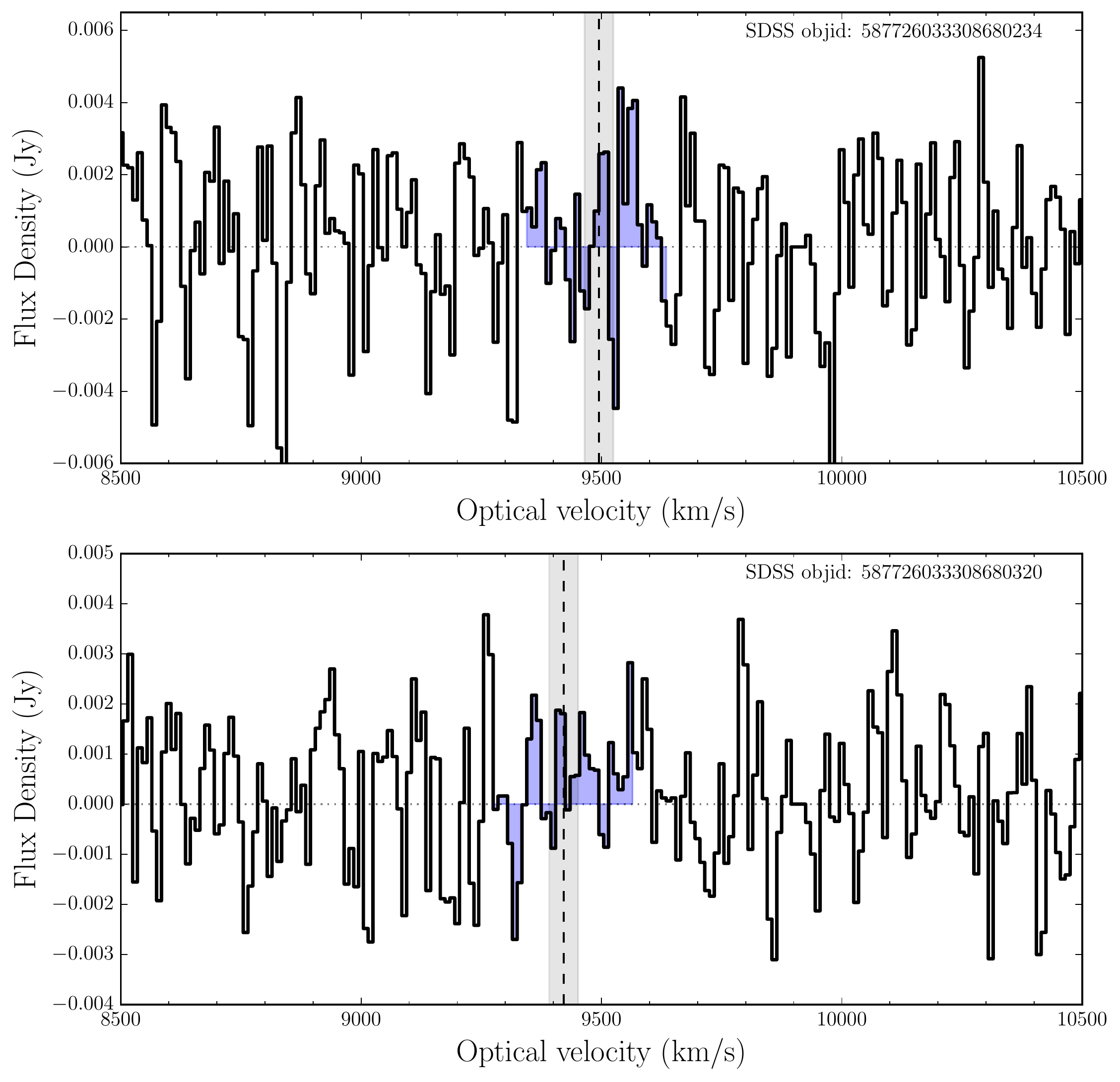}
    \end{minipage}
    \caption{Same as in Figure \ref{fig:0875}.  Left panel: SDSS thumbnail of galaxy pair 587726033308680234 (lower) \& 587726033308680320 (upper).
    Right panel: Flux density vs. optical velocity for galaxy pair 587726033308680234 (upper) \& 587726033308680320 (lower). The upper spectrum has a peak/RMS S/N of 1.97, while the lower spectrum has a peak/RMS S/N of 2.00.
    \label{fig:62963}}
\end{figure*}

\begin{figure*}
   \centering
   \begin{minipage}[c]{8.5cm}
   \includegraphics[height=245px]{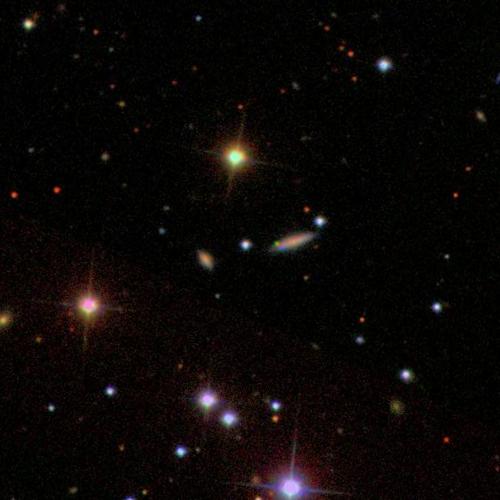} 
   \end{minipage}
   \begin{minipage}[c]{8.5cm}
   \includegraphics[height=250px]{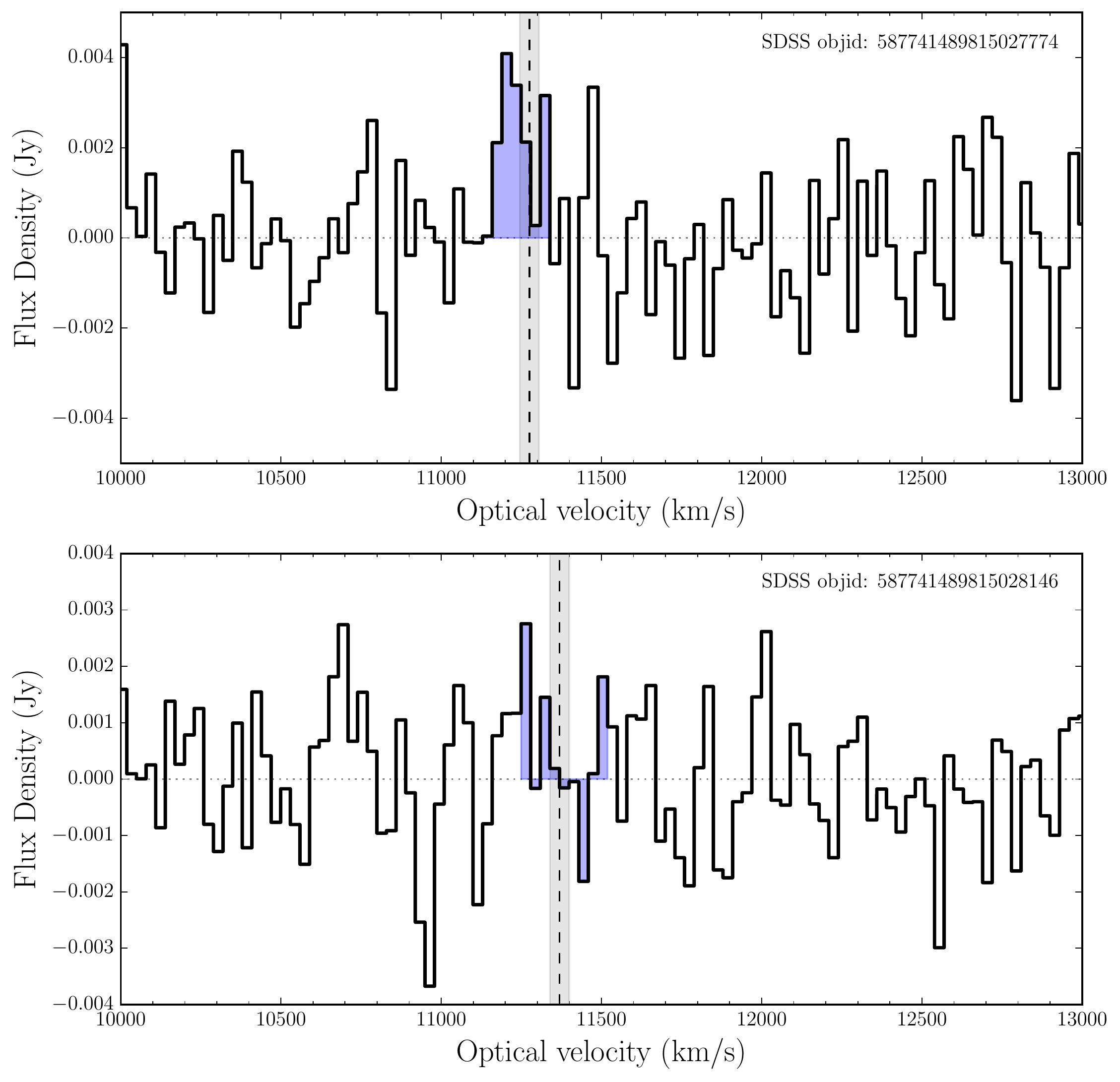}
    \end{minipage}
    \caption{Same as in Figure \ref{fig:0875}.  Left panel: SDSS thumbnail of galaxy pair 587741489815027774 (right) \& 587741489815028146 (left).
    Right panel: Flux density vs. optical velocity for galaxy pair 587741489815027774 (upper) \& 587741489815028146 (lower). The upper spectrum has a peak/RMS S/N of 2.70, while the lower spectrum has a peak/RMS S/N of 2.23.
    \label{fig:78236}}
\end{figure*}

\begin{figure*}
   \centering
   \begin{minipage}[c]{8.5cm}
   \includegraphics[height=245px]{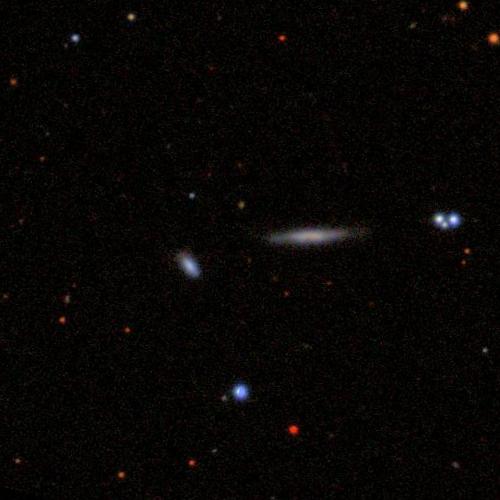} 
   \end{minipage}
   \begin{minipage}[c]{8.5cm}
   \includegraphics[height=250px]{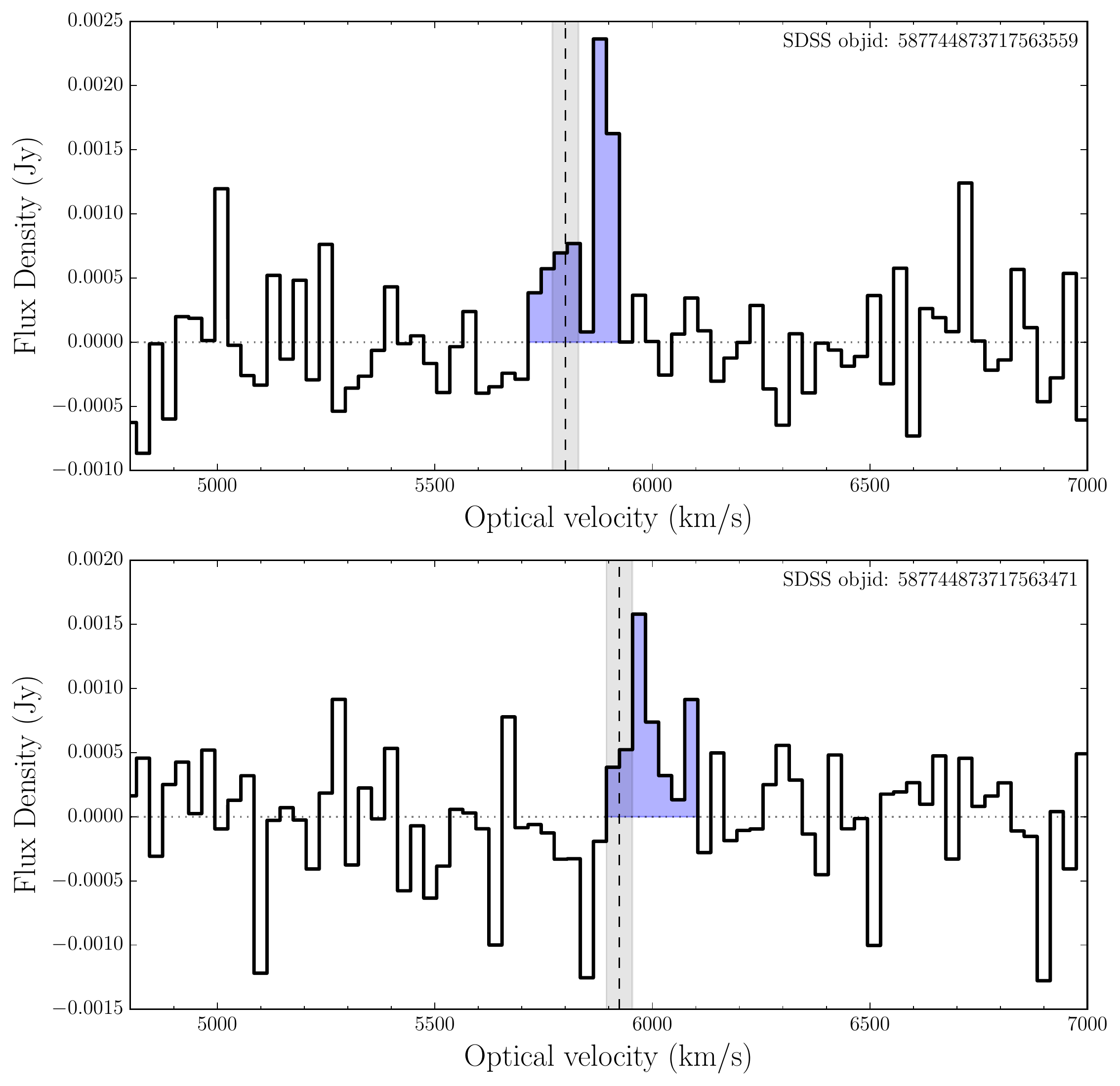}
    \end{minipage}
    \caption{Same as in Figure \ref{fig:0875}.  Left panel: SDSS thumbnail of galaxy pair 587744873717563559 (right) \& 587744873717563471 (left).
    Right panel: Flux density vs. optical velocity for galaxy pair 587744873717563559 (upper) \& 587744873717563471 (lower). The upper spectrum has a peak/RMS S/N of 5.35, while the lower spectrum has a peak/RMS S/N of 3.51.
    \label{fig:84028}}
\end{figure*}

\begin{figure*}
   \centering
   \begin{minipage}[c]{8.5cm}
   \includegraphics[height=245px]{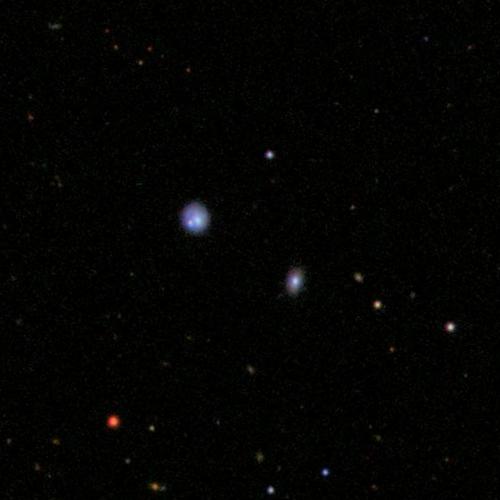} 
   \end{minipage}
   \begin{minipage}[c]{8.5cm}
   \includegraphics[height=250px]{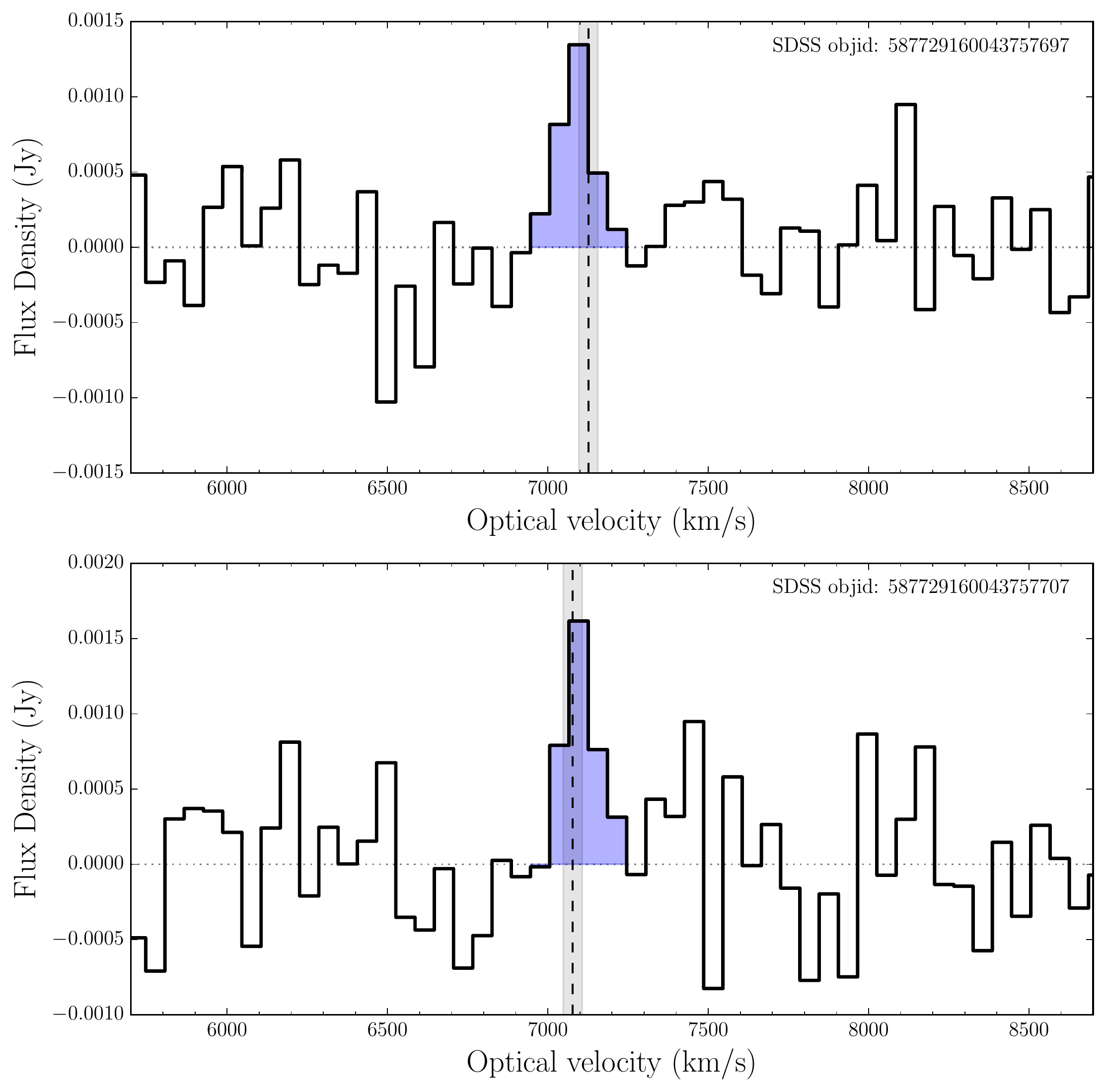}
    \end{minipage}
    \caption{Same as in Figure \ref{fig:0875}.  Left panel: SDSS thumbnail of galaxy pair 587729160043757697 (lower right) \& 587729160043757707 (upper left).
    Right panel: Flux density vs. optical velocity for galaxy pair 587729160043757697 (upper) \& 587729160043757707 (lower). The upper spectrum has a peak/RMS S/N of 3.76, while the lower spectrum has a peak/RMS S/N of 3.16.
    \label{fig:88885}}
\end{figure*}

\begin{figure*}
   \centering
   \begin{minipage}[c]{8.5cm}
   \includegraphics[height=245px]{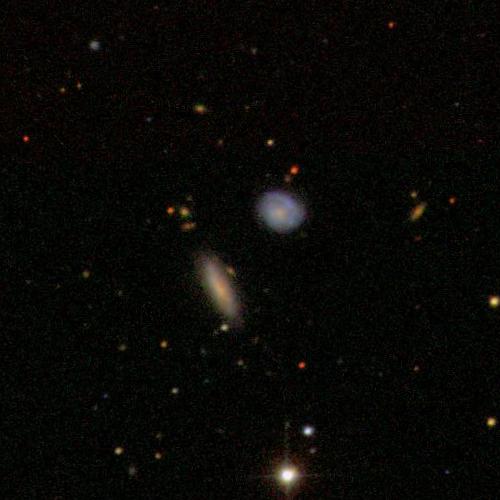} 
   \end{minipage}
   \begin{minipage}[c]{8.5cm}
   \includegraphics[height=250px]{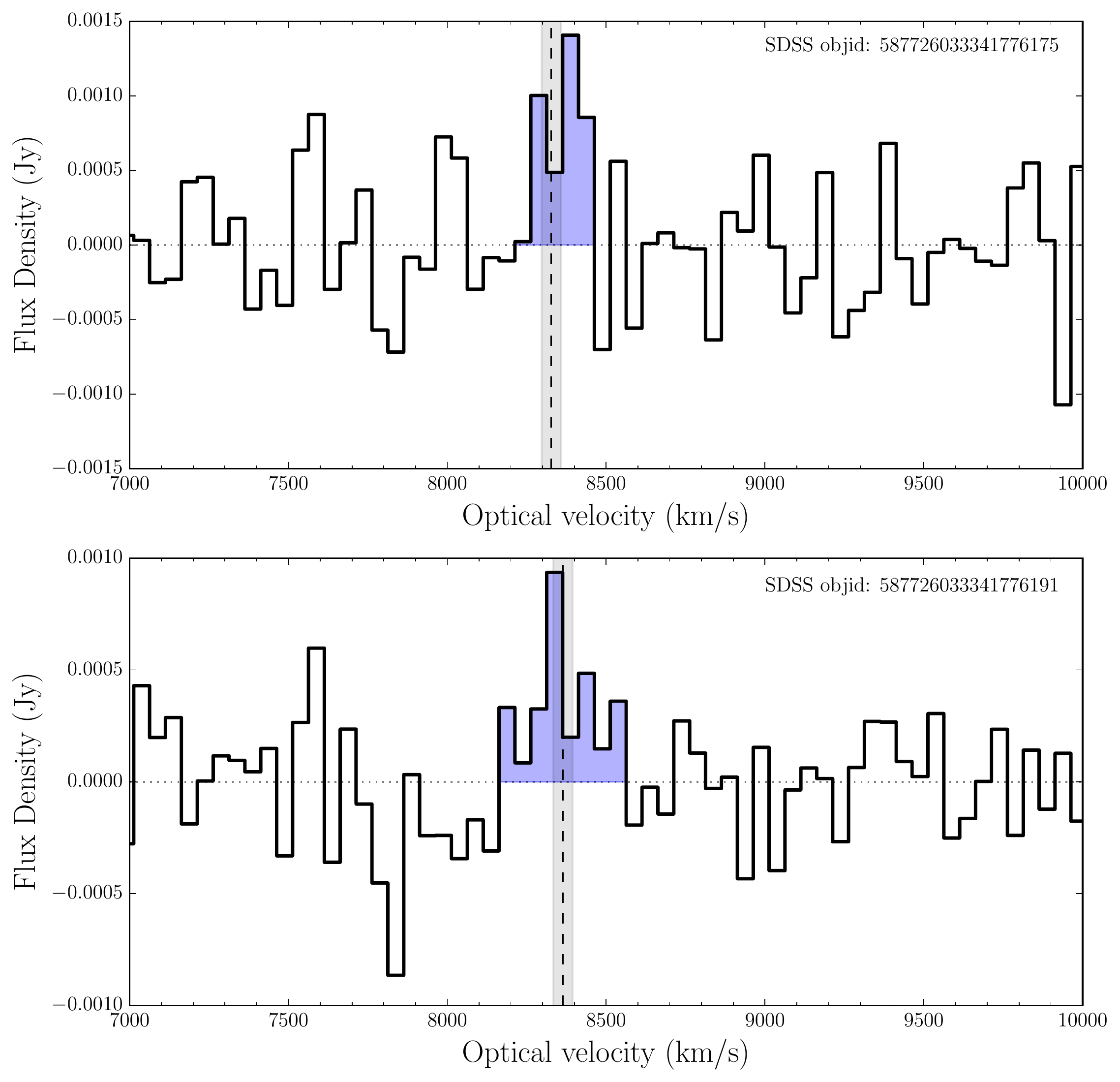}
    \end{minipage}
    \caption{Same as in Figure \ref{fig:0875}.  Left panel: SDSS thumbnail of galaxy pair 587726033341776175 (upper right) \& 587726033341776191 (lower left).
    Right panel: Flux density vs. optical velocity for galaxy pair 587726033341776175 (upper) \& 587726033341776191 (lower). The upper spectrum has a peak/RMS S/N of 2.85, while the lower spectrum has a peak/RMS S/N of 3.42.
    \label{fig:402775}}
\end{figure*}

In the Appendix, we present figures displaying the full data set.
Figure \ref{fig:UL_gf_tsfr} shows the full sample of 34 galaxies in our sample, including all non-detections.  
Figures \ref{fig:0875} -- \ref{fig:402775} show the SDSS image of the galaxy pair, along with the associated \hi~spectrum for each of the galaxies in the sample.  All images are at the same angular scale for consistency, and are 250 arcsec to a side.

\clearpage

\end{document}